\UseRawInputEncoding 

\documentclass[preprintnumbers, showpacs, floatfix,onecolumn,preprintnumbers, letterpaper, superscriptaddress,nofootinbib]{revtex4}
\usepackage{amsfonts}
\usepackage{amsmath}
\usepackage{amssymb,epsf}
\usepackage{latexsym}
\usepackage{graphicx,epsfig}
\usepackage{amssymb}
\usepackage{float}
\usepackage{subfigure}
\usepackage{epstopdf}
\usepackage[colorlinks=true,citecolor=blue,linkcolor=blue,urlcolor=black]{hyperref}
\usepackage{dcolumn}
\usepackage{psfrag}
\usepackage{wrapfig}
\usepackage{makeidx}
\usepackage{epsf}

\begin{document}

\title{Gauss-Bonnet holographic superconductors in lower-dimensions}
\author{Mahya Mohammadi}
\email{mahya689mohammadi@gmail.com}
 \affiliation{Department of Physics,
College of Sciences, Shiraz University, Shiraz 71454, Iran}
\affiliation{Biruni Observatory, College of Sciences, Shiraz
University, Shiraz 71454, Iran}

\author{Ahmad Sheykhi}
\email{asheykhi@shirazu.ac.ir}

 \affiliation{Department of Physics,
College of Sciences, Shiraz University, Shiraz 71454, Iran}
\affiliation{Biruni Observatory, College of Sciences, Shiraz
University, Shiraz 71454, Iran}

\begin{abstract}
We disclose the effects of Gauss-Bonnet gravity on the properties
of holographic $s$-wave and $p$-wave superconductors, with higher
order corrections, in lower-dimensional spacetime. We employ
shooting method to solve equations of motion numerically and
obtain the effect of different values of mass, nonlinear gauge
field and Gauss-Bonnet parameters on the critical temperature and
condensation. Based on our results, increasing each of these three
parameters leads to lower temperatures and larger values of
condensation. This phenomenon is rooted in the fact that
conductor/superconductor phase transition faces with difficulty
for higher effects of nonlinear and Gauss-Bonnet terms in the
presence of a massive field. In addition, we study the electrical
conductivity in holographic setup. In four dimension, real and
imaginary parts of conductivity in holographic $s$- and $p$-wave
models behave similarly and follow the same trend as higher
dimensions by showing the delta function and divergence behavior
at low frequency regime that Kramers-Kronig relation can connect
these two parts of conductivity to each other. We observe the
appearance of a gap energy at $\omega_{g}\approx8T_{c}$ which
shifts toward higher frequencies by diminishing temperature and
increasing the effect of nonlinear and Gauss-Bonnet terms.
Conductivity in three dimensions is far different from other
dimensions. Even the real and imaginary parts in $s$- and $p$-wave
modes pursue various trends. For example in $\omega\rightarrow 0$
limit, imaginary part in holographic $s$-wave model tends to
infinity but in $p$-wave model approaches to zero. However, the
real parts in both models show a delta function behavior. In
general, real and imaginary parts of conductivity in all cases
that we study tend to a constant value in $\omega\rightarrow
\infty$ regime.
\end{abstract}
\maketitle

\section{introduction}

The idea of holographic superconductors is introduced as a theoretical approach to study high temperature superconductors by applying AdS/CFT duality which indicates that the strong coupling
conformal field theory in $d$-dimension corresponds to a weak gravitational system in $(d+1)$-dimension \cite{Maldacena,H08,G98,W98,HR08,R10,H11,H09}. Since holographic superconductors(HSCs) show similar behavior as real superconductors, many works are done in this field such as considering the effect of nonlinear electrodynamics, different gravity models and backreaction analytically as well as numerically \cite{25,hendi,log,SSh16,Hg09,Gu09,HHH08,JCH10,SSh16,SH16,cai15,SHsh(17),
Ge10,Ge12,Kuang13,Pan11,CAI11, SHSH(16),shSh(16),Doa, Afsoon,
cai10,yao13,n4,n5,n6,Gan1,mahya}. Like real superconductors, we can classify HSCs to $s$-, $p$- and $d$-wave models\cite{BCS57,superp,Caip,cai13p,Donos,Gubser,chaturverdip15,Roberts8,zeng11,cai11p,pando12,momeni12p,mahyap,francessco1,francessco2,mahyalif}.

Gauss-Bonnet gravity as a higher order correction in the phase transition of gravity model has been studied in $D\geq 5$ for a long time. 
However, it has been found that this theory can be considered in lower dimensions by applying some techniques. According to \cite{4gb1}, rescaling the Gauss-Bonnet gravity term $\alpha\rightarrow \alpha/(D-4)$ and taking the limit $D\rightarrow 4$ causes singularity resolution and makes study the higher order corrections in $4$-dimensional possible\cite{4gb1}. By following this approach, many researches are done to explore different aspects of Gauss-Bonnet gravity in $D=4$ such as gravitational lensing, the thermodynamics, the observable shadows and the quasinormal modes \cite{4gb2,4gb3,4gb4,4gb5,4gb6,4gb7,4gb8,4gb9,4gb10,4gb11}.
However, some criticisms on this method were revealed and argued that there is no pure four-dimensional Gauss-Bonnet gravity\cite{4gb12,4gb13,4gb14,4gb15}. In \cite{RBM1,RBM2}, a new method to study lower dimensional Gauss-Bonnet gravity was presented. Studying holographic superconductors in the presence of Gauss-Bonnet correction in $D=4$ by using the rescaling the Gauss-Bonnet term are done numerically and analytically \cite{hscgb4,hscgb42}.

The purpose of this work is to explore the properties of holographic $s$- and $p$-waves
superconductors in $D=3$ and $D=4$ dimensions by taking into account
the higher order corrections on the gravity and gauge field sides. We consider the effect of Gauss-Bonnet gravity in lower dimensions according to \cite{RBM1,RBM2}. Moreover, We apply general nonlinear electrodynamics with
higher order correction term, namely $\mathcal{L}=\mathcal{F}+b
\mathcal{F}^{2}$, on the gauge field side. There are various ways to analyze holographic $p$-wave superconductors but in the present work, we describe this kind of HSCs by considering a vector field in the bulk which
corresponds to the vector order parameter in the boundary\cite{Caip,cai13p,Donos,Gubser,chaturverdip15}.

We solve equations of motion numerically by applying shooting method to seek the effect of mass, nonlinear and Gauss-Bonnet coefficients on critical temperature and condensation. In all cases, increasing these three parameters diminishes the critical temperature while condensation values go up. This behavior is a consequence of the fact that scalar/vector hair forms difficult in the presence of strong nonlinear and Gauss-Bonnet effects for a massive scalar/vector field. The trend of condensation as a function of temperature is the same as in BCS theory by following the mean field theory.  Then we obtain the electrical conductivity. In order to understand the influence of mass, nonlinear and Gauss-Bonnet terms, we plot
the real and imaginary parts of
conductivity as a function of frequency. We get that both parts of conductivity in $D=4$ are identical in both models and follow the same approach as higher orders. A good illustration of this is obeying the Kramers-kronig relation by having a delta function and a pole in real and imaginary parts of conductivity. In addition, the gap frequency occurs below the critical temperature at about $\omega_{g}/T_{c}\simeq 8$ which shifts to higher frequencies by decreasing temperature or increasing the effect of nonlinear and Gauss-Bonnet terms. However, real and imaginary parts of conductivity in $D=3$ behave far different from each other and other dimensions. Maybe, considering the effect of backreaction can improve the results of conductivity in this dimension.

This work is outlined as follows. In section \ref{sec2}, we
introduce the holographic setup in the presence of Gauss-Bonnet gravity. In section \ref{sections}, we present the behavior of condensation and conductivity of holographic $s$-wave superconductors in $D=3$ and $D=4$. Section \ref{sectionp} is devoted to study conductor/superconductor phase transition and conductivity of holographic $p$-wave models in $D=3$ and $D=4$. Finally, we summarize our results in
section \ref{section5}.

\section{Holographic setup}\label{sec2}
We can express the action of gravitational field in the presence of Gauss-Bonnet gravity in lower dimensions as \cite{RBM1}
\begin{eqnarray}
&&S_{G} =\int d^{D}x\sqrt{-g} \left[R - 2 \Lambda +\alpha \left(\zeta \mathcal{G} + 4 G^{\mu \nu}\partial_{\mu} \zeta \partial_{\nu} \zeta\ - 4(\partial\zeta)^{2}\square \zeta+2((\partial\zeta)^{2})^{2} \right)\right], \notag \\
&& \mathcal{G}=R_{\mu \nu \rho \sigma} R^{\mu \nu \rho \sigma}- 4 R_{\mu \nu} R^{\mu \nu}+ R^{2}, \notag \\
\label{actgauss}
\end{eqnarray}%
where $g$ and $R$ are metric determinant and Ricci scalar, respectively.
 In addition, by introducing $l$ as the radius of the AdS spacetime, the negative cosmological constant can be defined by \cite{mahyagaussp}
\begin{equation}
\Lambda=-\frac{(d-1)(d-2)}{2 l^2}.
\end{equation}
Furthermore, Ricci tensor and Riemann curvature tensor are defined by $\alpha$, $R_{\mu \nu}$ and $R_{\mu \nu \rho \sigma}$.
In this model $\zeta$ represents a scalar field. If in equation(\ref{actgauss}) the Gauss-Bonnet parameter tends to zero, $\alpha\rightarrow0$, the action in Einstein case is recovered. We can define the metric as follows \cite{RBM1}
\begin{eqnarray} \label{metric2}
&&{ds}^{2}=-f(r){dt}^{2}+\frac{{dr}^{2}}{f(r)}+r^{2} \sum _{i=1}^{D-2}{dx_{i}}^{2}%
,\\
&&f(r)=\frac{r^2}{2 \alpha } \left[1-\sqrt{1-4 \alpha \left(1-\frac{1}{r^{(D-1)}}\right)}\right],\label{eqfgauss} %
\end{eqnarray}%
where the function $f(r)$ has the asymptotic behavior as
\begin{equation}
f(r)=\frac{ r^2}{2 \alpha }\left[1-\sqrt{1-4 \alpha }\right].
\end{equation}

We can present the effective radius $L_{\text{eff}}$ for the AdS spacetime as \cite{RBM1}
\begin{equation}
L_{\text{eff}}^2=\frac{2 \alpha }{1-\sqrt{1-4 \alpha }}.
\end{equation}
Based on the above equation, in order to have a well-defined vacuum expectation value $\alpha \leq 1/4$ where the upper bound $\alpha = 1/4$ is called Chern-Simon limit \cite{caipp}.

The Hawking temperature of the black hole is given by
\begin{equation}
T=\frac{f^{^{\prime }}(r_{+})}{4\pi }=\frac{(D-1) r_{+}}{4\pi}, \label{temp}
\end{equation}%
where $r_{+}$ represents the black hole horizon radius.
\section{Holographic $s$-wave superconductors}\label{sections}
\section*{Condensation of scalar field}\label{sec3}
\begin{figure*}[t]
\centering
\subfigure[~$\overline{m}^{2}=-3/4$]{\includegraphics[width=0.4\textwidth]{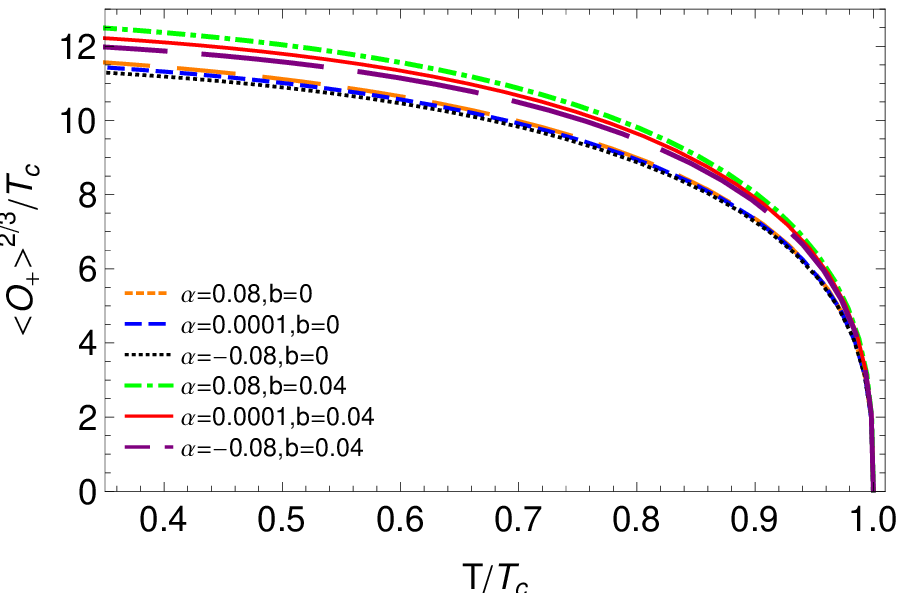}} \qquad %
\subfigure[~$\overline{m}^{2}=0$]{\includegraphics[width=0.4\textwidth]{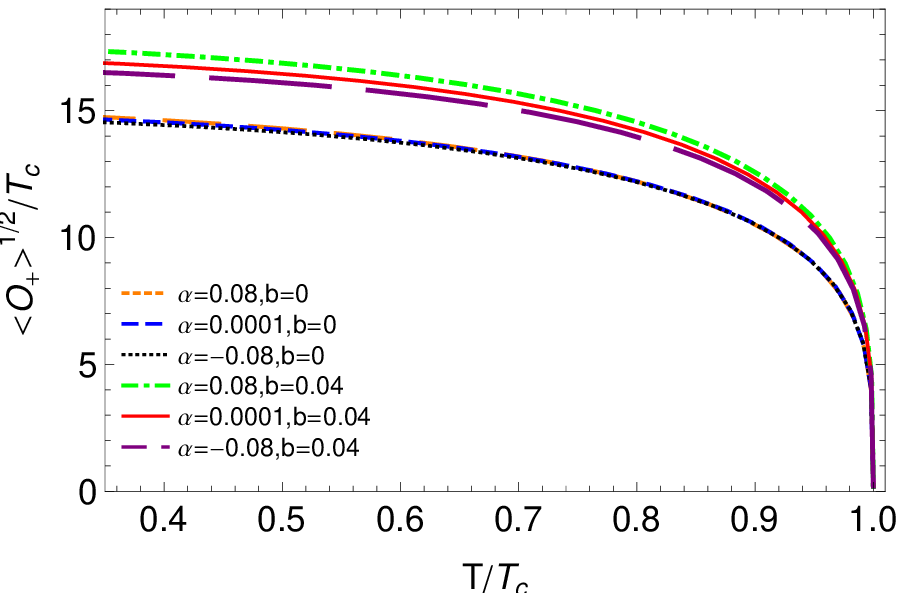}} \qquad %
\caption{The behavior of the condensation parameter as a function
of the temperature for different values of nonlinearity parameters
in $D=3$ in $s$-wave case.} \label{fig1}
\end{figure*}

\begin{figure*}[t]
\centering
\subfigure[~$\overline{m}^{2}=-5/4$]{\includegraphics[width=0.4\textwidth]{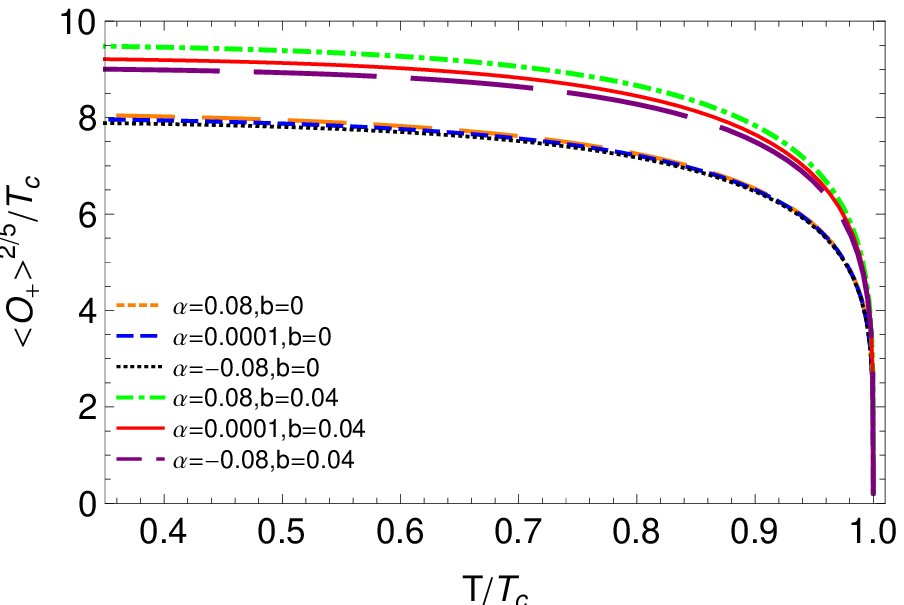}} \qquad %
\subfigure[~$\overline{m}^{2}=0$]{\includegraphics[width=0.4\textwidth]{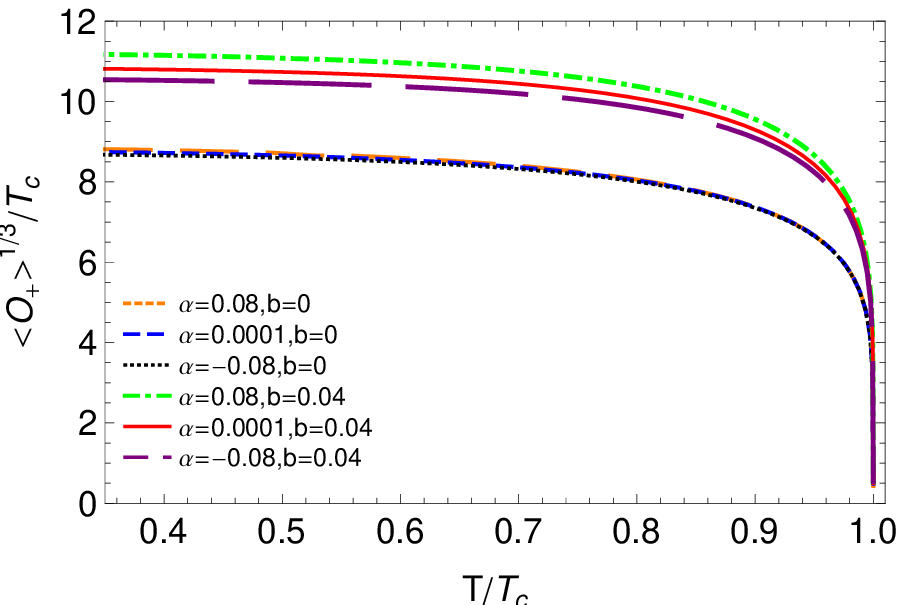}} \qquad %
\caption{The behavior of the condensation parameter as a function
of the temperature for different values of nonlinearity parameters
in $D=4$ in $s$-wave case.} \label{fig2}
\end{figure*}

To describe a holographic $s$-wave superconductor by considering $m$ and $q$ as mass and charge of scalar field $\psi$, action of matter part has the following form
\begin{equation}\label{actms}
S_{m} =\int d^{D}x\sqrt{-g} \left[
\mathcal{L}_{\mathcal{NL}}- \vert\bigtriangledown\psi-i q A \psi \vert^{2}
-m^{2} \vert \psi \vert^{2}\right],
\end{equation}
$\mathcal{L}_{\mathcal{NL}}$ represents the Lagrangian
density of nonlinear electrodynamics which defines as
\begin{equation}\label{eqnon}
\mathcal{L}_{\mathcal{NL}}=\mathcal{F}+b \mathcal{F}^{2}, \ \ \ \ \mathcal{F}=-\frac{1}{4} F_{\mu\nu}F^{\mu\nu},
\end{equation}%
when the nonlinearity parameter $b$ tends to zero, $\mathcal{L}_{\mathcal{NL}}$
turns to the standard Maxwell Lagrangian $-1/4 F_{\mu\nu}F^{\mu\nu}$.
The strength of the Maxwell field equals to $F_{\mu\nu}=\nabla_{\mu} A_{\nu}-\nabla_{\nu} A_{\mu}$ with $A_{\mu}$ as the vector
potential.
After varying the action with respect to $A_{\mu}$ and $\psi$ and adopting the ansatz for the gauge and the scalar fields as $A_{t}=\phi(r)$ and $\psi=\psi(r)$, the equations of motion are obtained as
\begin{equation}\label{phis}
\phi ''(r)+\frac{\phi '(r) (D-2)}{r} \left[\frac{b \phi '(r)^2+1}{3 b \phi '(r)^2+1}\right]-\frac{2 q^2 \psi (r)^2 \phi (r)}{f(r)}\left[\frac{1}{ 3 b \phi '(r)^2+1}\right]=0,
\end{equation}
\begin{equation}\label{psis}
\psi ''(r)+\psi '(r)\left[\frac{f'(r)}{f(r)}+\frac{(D-2)}{r}\right] +\psi (r) \left[\frac{q^2 \phi (r)^2}{f(r)^2}-\frac{m^2}{f(r)}\right]=0,
\end{equation}
in the Maxwell limit, $b\rightarrow0$, field equations turn to corresponding equations in \cite{H08} with $D=4$. At the vicinity of AdS boundary($r\rightarrow \infty$), these equations have the asymptotic behavior as
\begin{equation} \label{asymphis}
\phi(r) = \left\{
\begin{array}{lr}
\rho +\mu ln(r), & D=3\\
\bigskip\\
\mu-\dfrac{\rho}{r}, & D=4\\
\end{array} \right.
\end{equation}%
\begin{equation}\label{asympsis}
\psi(r)=\dfrac{\psi_{+}(r)}{r^{\Delta_{+}}}+\dfrac{\psi_{-}(r)}{r^{\Delta_{-}}}, \ \ \ \ \Delta _\pm=\frac{1}{2} \left[(D-1)\pm\sqrt{(D-1)^2+4 m^2 L_{\text{eff}}^{2}}\right],
\end{equation}
based on the above equation, our choosing masses should satisfy the Breitenlohner-Freedman (BF) bound as \cite{wen18}
\begin{equation}\label{bfs}
\overline{m}^{2}\geqslant - \frac{(D-1)^{2}}{4}, \ \ \ \ \overline{m}^{2}=m^{2} L_{\text{eff}}^{2}.
\end{equation}
where $\mu$ and $\rho$ are interpreted as chemical potential and charge density in the dual field theory. In addition based on the AdS/CFT duality, $\psi_{+}$ and $\psi_{-}$ are regarded as duals of expectation value of order parameter and its source. Since we are looking for spontaneous symmetry breaking, we give the role of source to $\psi_{-}$ and set it to zero. $\psi_{+}$ is considered as expectation value of the order parameter $\langle O_{+} \rangle$ in this work. To pursue our research, we solved equations (\ref{phis}) and (\ref{psis}) numerically by applying shooting method. Without loss of generality, we set $r_{+}=q=1$. Our numerical results of critical temperatures $T_{c}$ based on $\mu$ or $\sqrt{\rho}$ affected by different values of mass, nonlinearity and Gauss-Bonnet parameters
 are summarized in tables I and II. According to our results, increasing each of these three parameters makes the condensation harder to form by diminishing the critical temperature.
Figures \ref{fig1}-\ref{fig2}, show the behavior of condensation $\langle O_{+} \rangle^{1/(1+\Delta_{+})}$ as a function of temperature for different choices of mass, nonlinearity and Gauss-Bonnet parameters in $D=3$ and $D=4$. The condensation values raise by increasing each of theses three parameters. Besides, the Gauss-Bonnet term does not change the critical exponent of condensation which means that we face with second order phase transition for all values of $\alpha$. Increasing the value of nonlinearity parameters makes the effect of different Gauss-Bonnet terms more clear. We obtain the same results as \cite{liu} in $D=3$ with $\overline{m}^{2}=0$ and $\alpha=0.0001$ which proves that for small values of $\alpha$ the Einstein case is regained.
\begin{table*}[t]
\label{tab1}
\begin{center}
\begin{tabular}{c|c|c|c|c|c|c|}
\cline{2-3}\cline{2-7}\cline{4-7}
& \multicolumn{2}{|c|}{$b=0$} & \multicolumn{2}{|c|}{$b=0.02$} &
\multicolumn{2}{|c|}{$b=0.04$} \\ \cline{2-3}\cline{2-7}\cline{4-7}
& $\overline{m}^{2}=-3/4$ & $\overline{m}^{2}=0$ & $\overline{m}^{2}=-3/4$ & $\overline{m}^{2}=0$ & $\overline{m}^{2}=-3/4$ & $\overline{m}^{2}=0$
\\ \hline
\multicolumn{1}{|c|}{$\alpha=0.08$} & $0.0652$ $\mu$ & $ 0.0425$  $\mu$ & $0.0621$  $\mu$ & $%
0.0380$  $\mu$ & $0.0596$  $\mu$ & $0.0348$  $\mu$ \\ \hline
\multicolumn{1}{|c|}{$\alpha=0.0001$} & $0.0704$  $\mu$ & $0.0460$  $\mu$ & $0.0675$  $\mu$ & $%
0.0418$  $\mu$ & $0.0651$  $\mu$ & $0.0386$  $\mu$ \\ \hline
\multicolumn{1}{|c|}{$\alpha=-0.08$} & $0.0849$  $\mu$ & $0.0490$  $\mu$ & $0.0722$  $\mu$ & $%
0.0450$  $\mu$ & $0.0699$ $\mu$ & $0.0419$  $\mu$ \\ \hline
\end{tabular}%
\caption{Numerical results for critical temperature $T_{c}$ for different values of mass, nonlinearity and Gauss-Bonnet parameters for holographic $s$-wave superconductor in $D=3$.}
\end{center}
\end{table*}

\begin{table*}[t]
\label{tab4}
\begin{center}
\begin{tabular}{c|c|c|c|c|c|c|}
\cline{2-3}\cline{2-7}\cline{4-7}
& \multicolumn{2}{|c|}{$b=0$} & \multicolumn{2}{|c|}{$b=0.02$} &
\multicolumn{2}{|c|}{$b=0.04$} \\ \cline{2-3}\cline{2-7}\cline{4-7}
& $\overline{m}^{2}=-5/4$ & $\overline{m}^{2}=0$ & $\overline{m}^{2}=-5/4$ & $\overline{m}^{2}=0$ & $\overline{m}^{2}=-5/4$ & $\overline{m}^{2}=0$
\\ \hline
\multicolumn{1}{|c|}{$\alpha=0.08$} & $0.0954$ $\sqrt{\rho}$ & $0.0833 $  $\sqrt{\rho}$ & $0.0853 $  $\sqrt{\rho}$ & $0.0700 $  $\sqrt{\rho}$ & $0.0793 $  $\sqrt{\rho}$ & $0.0633 $  $\sqrt{\rho}$ \\ \hline
\multicolumn{1}{|c|}{$\alpha=0.0001$} & $0.0992$  $\sqrt{\rho}$ & $0.0866 $  $\sqrt{\rho}$ & $0.0899 $  $\sqrt{\rho}$ & $0.0744 $  $\sqrt{\rho}$ & $ 0.0842$  $\sqrt{\rho}$ & $0.0678 $  $\sqrt{\rho}$ \\ \hline
\multicolumn{1}{|c|}{$\alpha=-0.08$} & $0.1024$  $\sqrt{\rho}$  & $0.0895 $  $\sqrt{\rho}$ & $0.0938 $  $\sqrt{\rho}$ & $%
0.0799$  $\sqrt{\rho}$ & $0.0883 $ $\sqrt{\rho}$ & $0.0715 $  $\sqrt{\rho}$ \\ \hline
\end{tabular}%
\caption{Numerical results for critical temperature $T_{c}$ for different values of mass, nonlinearity and Gauss-Bonnet parameters for holographic $s$-wave superconductor in $D=4$.}
\end{center}
\end{table*}

\section*{Conductivity}\label{sec4}
\begin{figure*}[t]
\centering
\subfigure[~$\alpha=0.08$]{\includegraphics[width=0.4\textwidth]{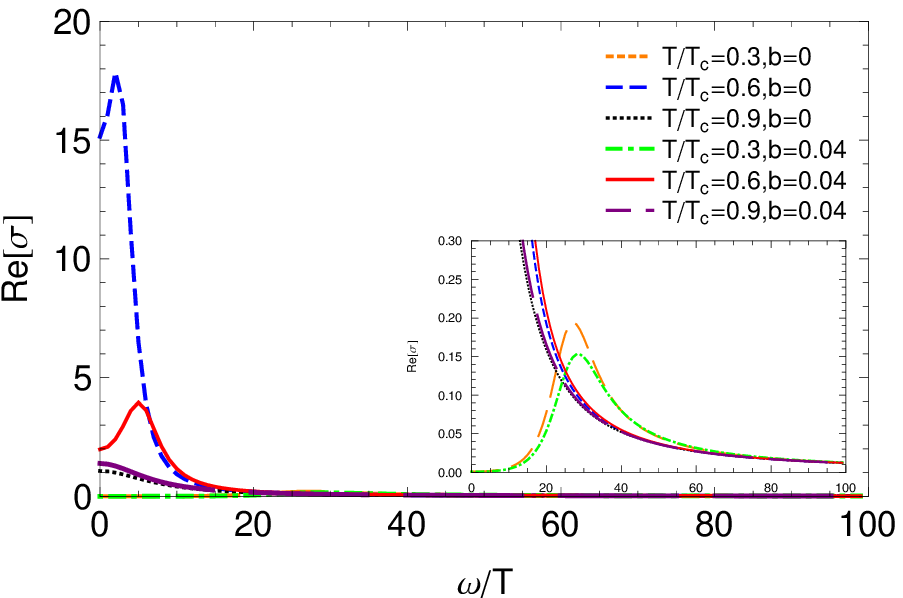}} \qquad %
\subfigure[~$\alpha=-0.08$]{\includegraphics[width=0.4\textwidth]{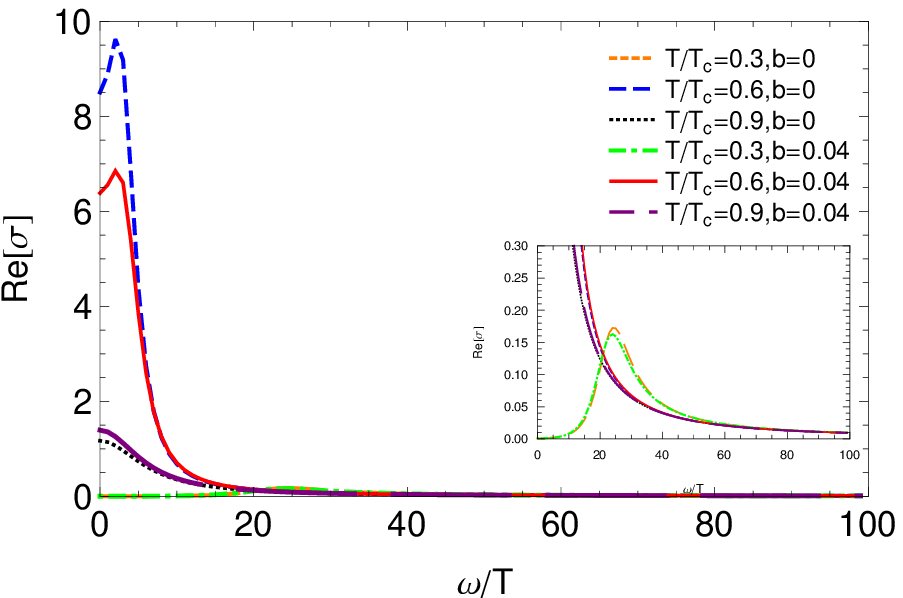}} \qquad %
\subfigure[~$\alpha=0.08$]{\includegraphics[width=0.4\textwidth]{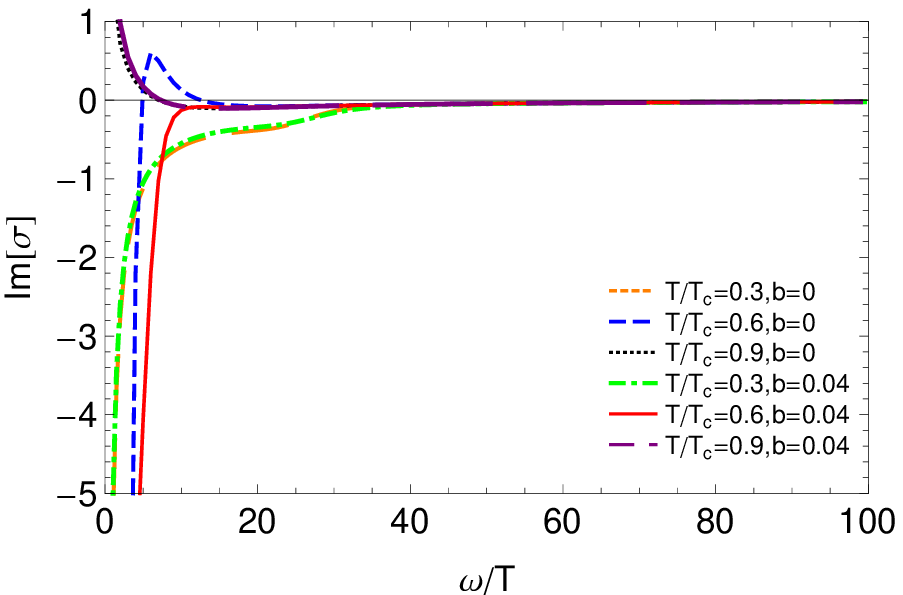}} \qquad %
\subfigure[~$\alpha=-0.08$]{\includegraphics[width=0.4\textwidth]{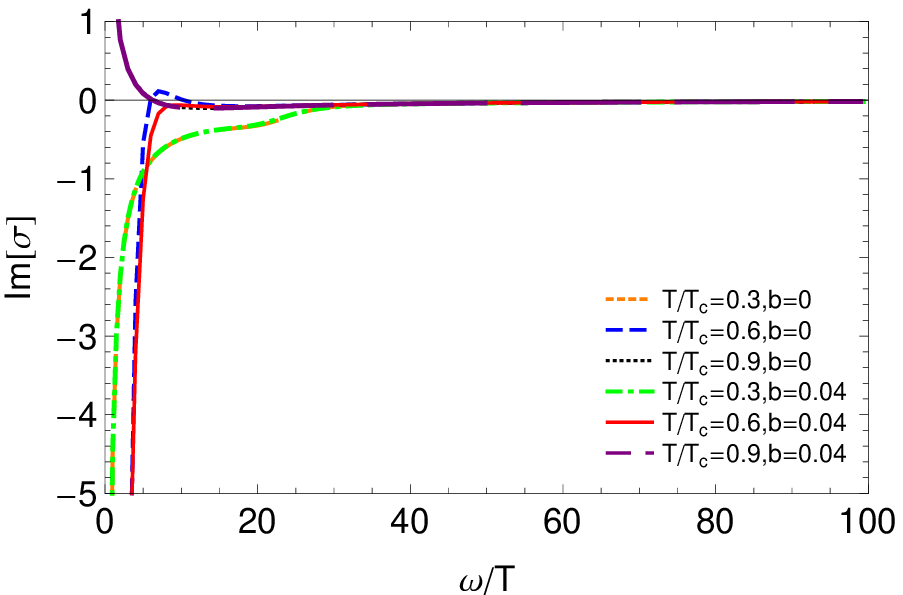}} \qquad %
\caption{The behavior of real and imaginary parts of conductivity with
$\overline{m}^{2}=-3/4$ in $D=3$ in $s$-wave case.} \label{fig3}
\end{figure*}

\begin{figure*}[t]
\centering
\subfigure[~$b=0$, $\alpha=0.08$]{\includegraphics[width=0.4\textwidth]{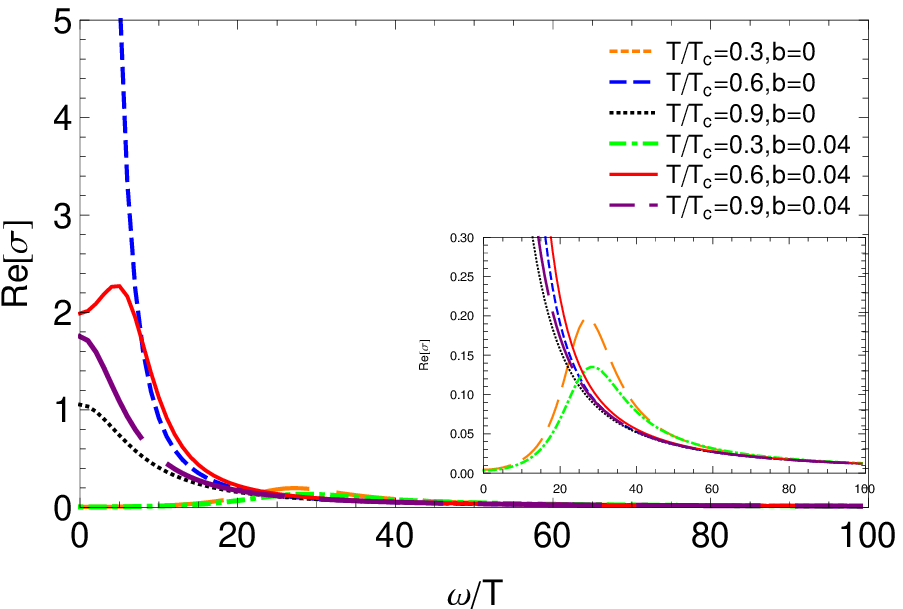}} \qquad %
\subfigure[~$\alpha=-0.08$]{\includegraphics[width=0.4\textwidth]{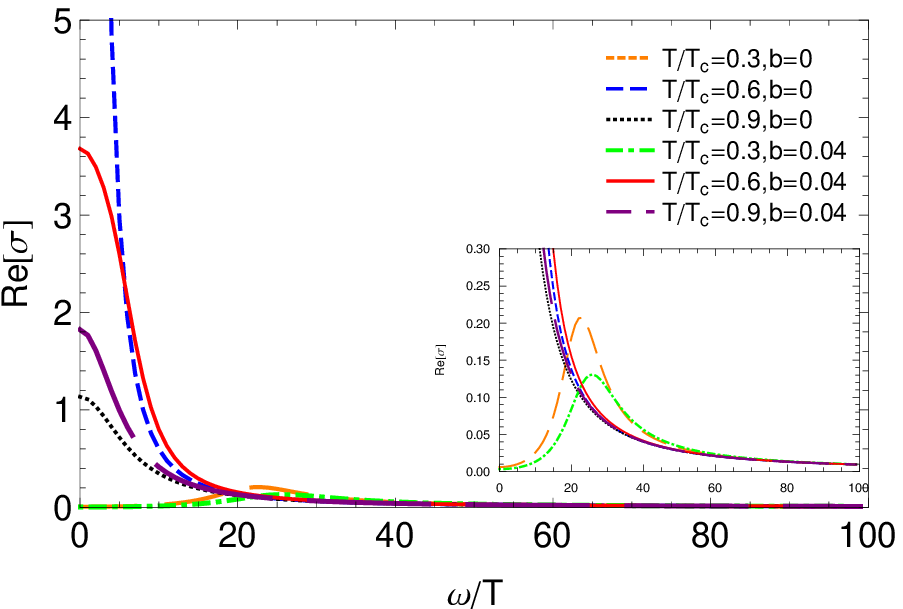}} \qquad %
\subfigure[~$\alpha=0.08$]{\includegraphics[width=0.4\textwidth]{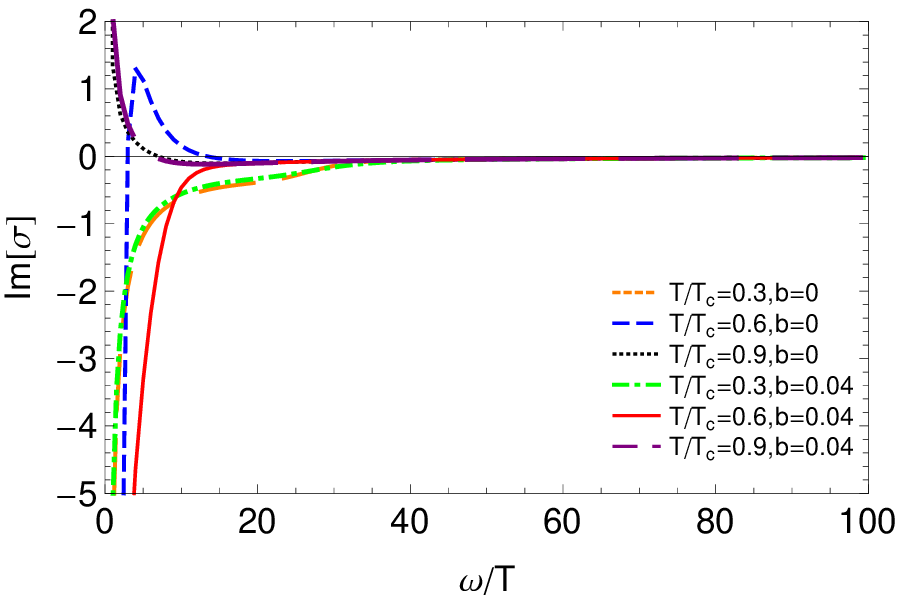}} \qquad %
\subfigure[~$\alpha=-0.08$]{\includegraphics[width=0.4\textwidth]{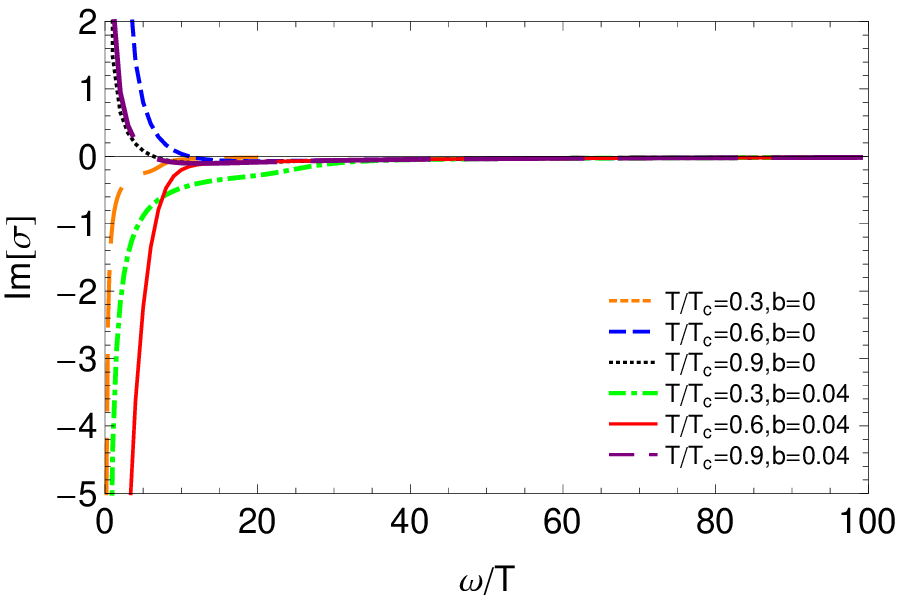}} \qquad %
\caption{The behavior of real and imaginary parts of conductivity with
$\overline{m}^{2}=0$ in $D=3$ in $s$-wave case.} \label{fig4}
\end{figure*}

\begin{figure*}[t]
\centering
\subfigure[~$b=0$, $\alpha=0.08$]{\includegraphics[width=0.4\textwidth]{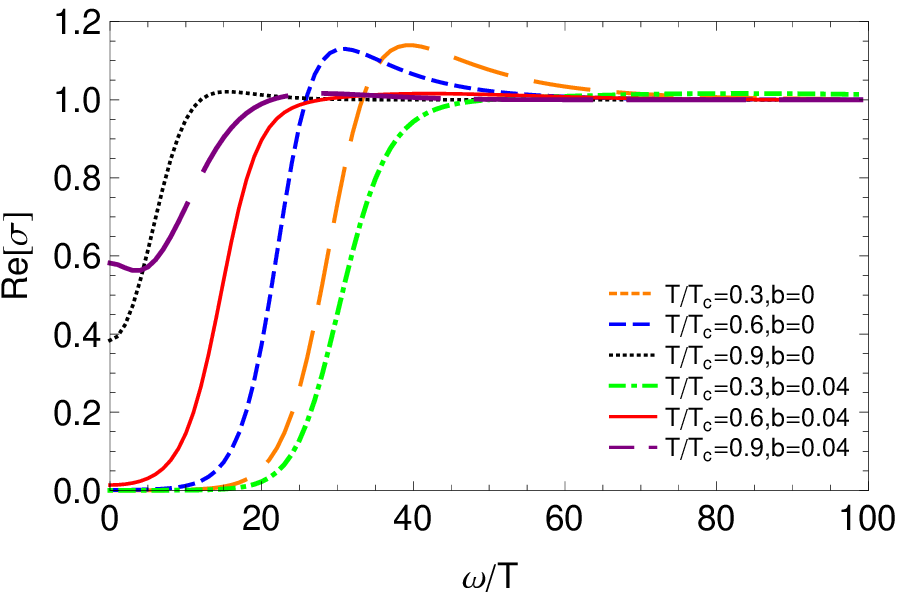}} \qquad %
\subfigure[~$\alpha=-0.08$]{\includegraphics[width=0.4\textwidth]{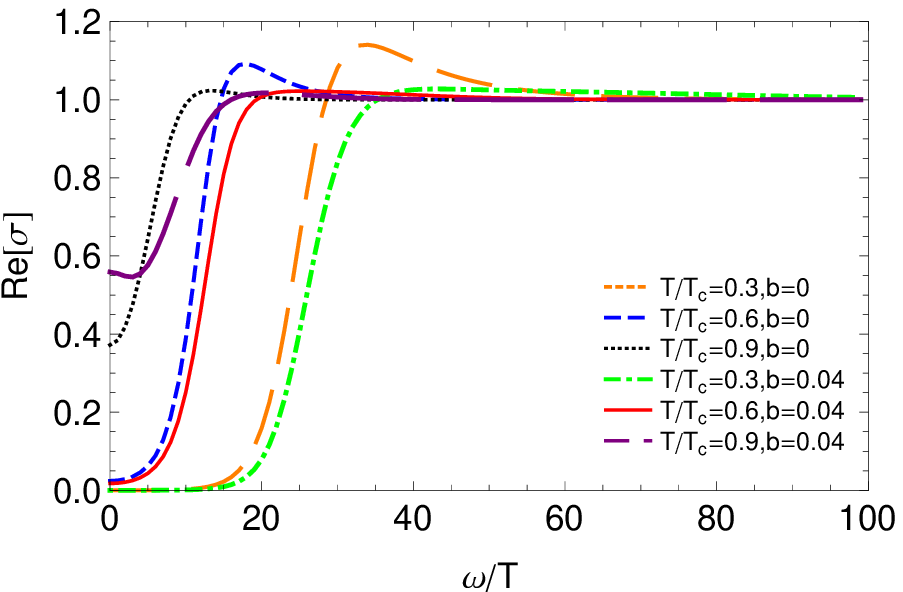}} \qquad %
\subfigure[~$\alpha=0.08$]{\includegraphics[width=0.4\textwidth]{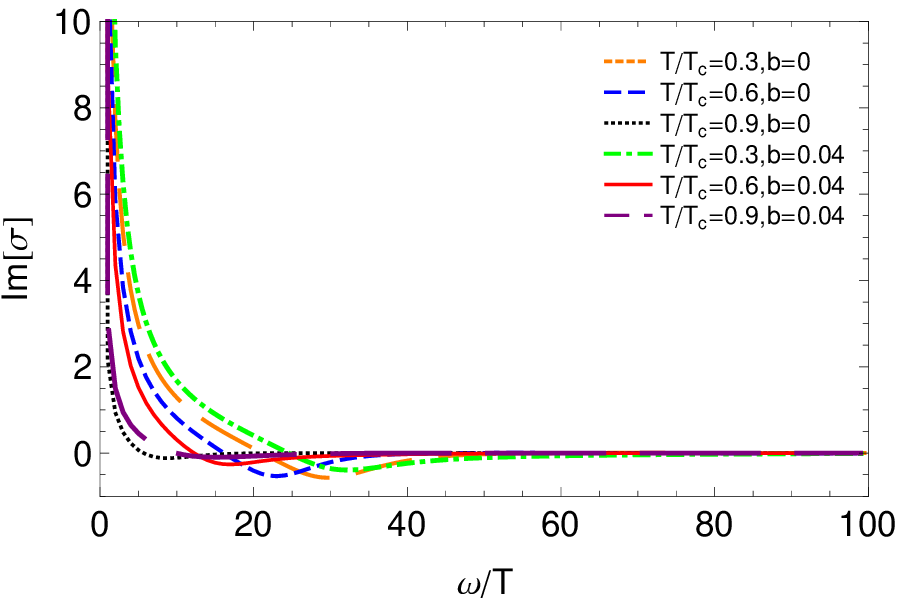}} \qquad %
\subfigure[~$\alpha=-0.08$]{\includegraphics[width=0.4\textwidth]{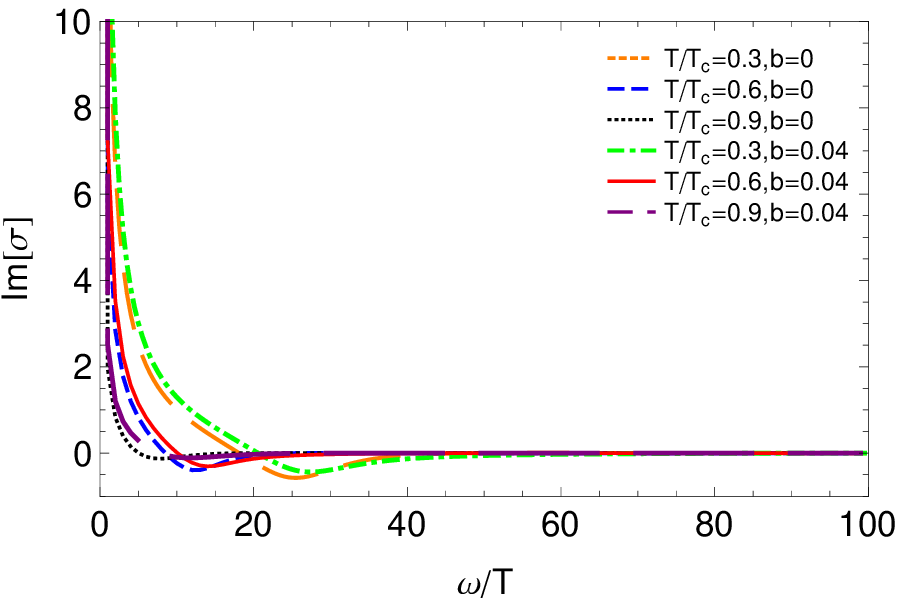}} \qquad %
\caption{The behavior of real and imaginary parts of conductivity with $\overline{m}^{2}=-5/4$ in $D=4$ in $s$-wave case.}
\label{fig5}
\end{figure*}

\begin{figure*}[t]
\centering
\subfigure[~$b=0$, $\alpha=0.08$]{\includegraphics[width=0.4\textwidth]{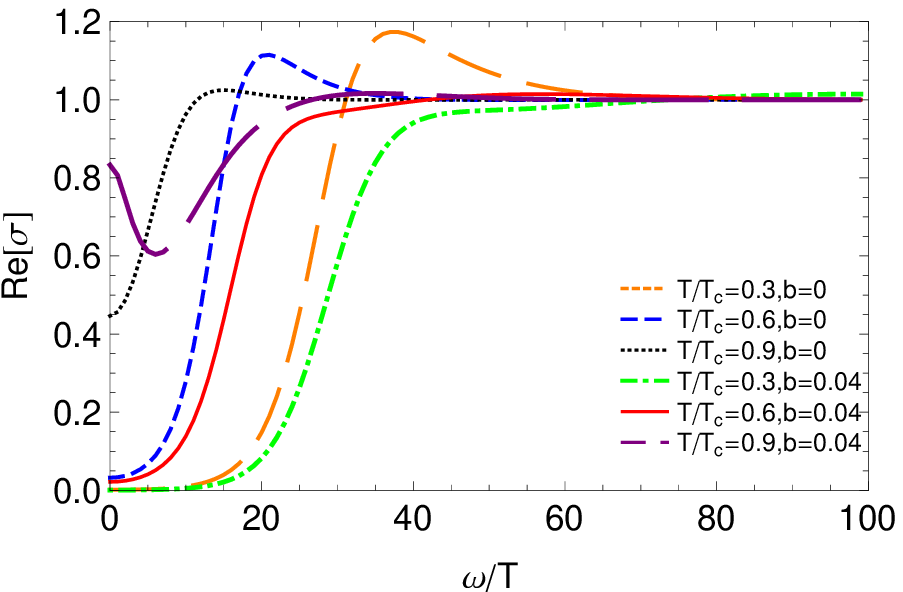}} \qquad %
\subfigure[~$\alpha=-0.08$]{\includegraphics[width=0.4\textwidth]{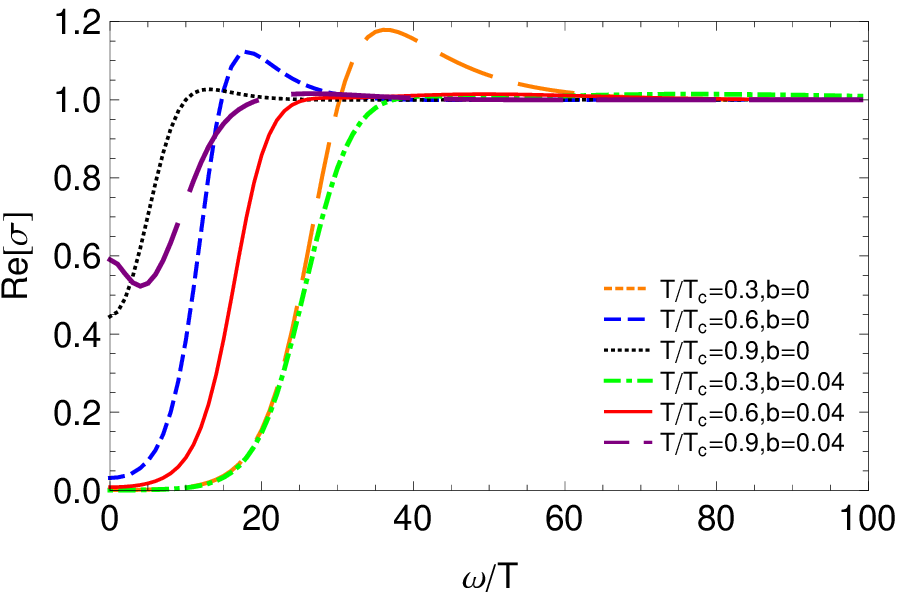}} \qquad %
\subfigure[~$\alpha=0.08$]{\includegraphics[width=0.4\textwidth]{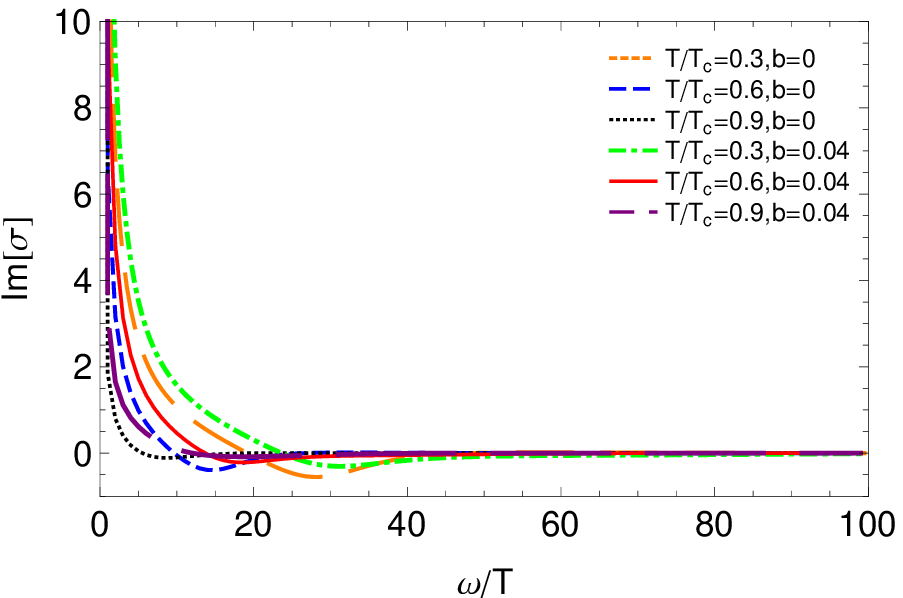}} \qquad %
\subfigure[~$\alpha=-0.08$]{\includegraphics[width=0.4\textwidth]{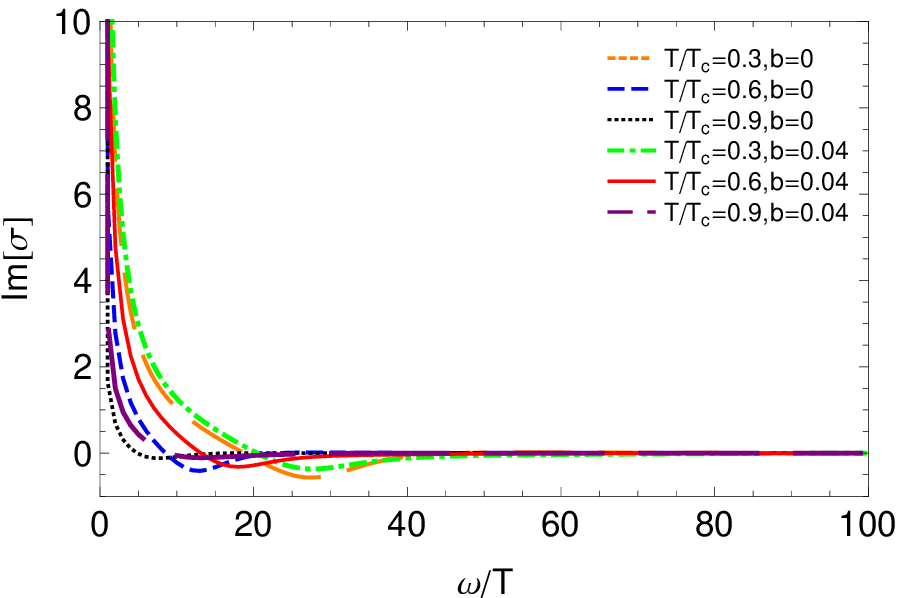}} \qquad %
\caption{The behavior of real and imaginary parts of conductivity with $\overline{m}^{2}=0$ in $D=4$ in $s$-wave case.}
\label{fig6}
\end{figure*}

\begin{figure*}[t]
\centering
\subfigure[~$\overline{m}^{2}=-3/4$]{\includegraphics[width=0.4\textwidth]{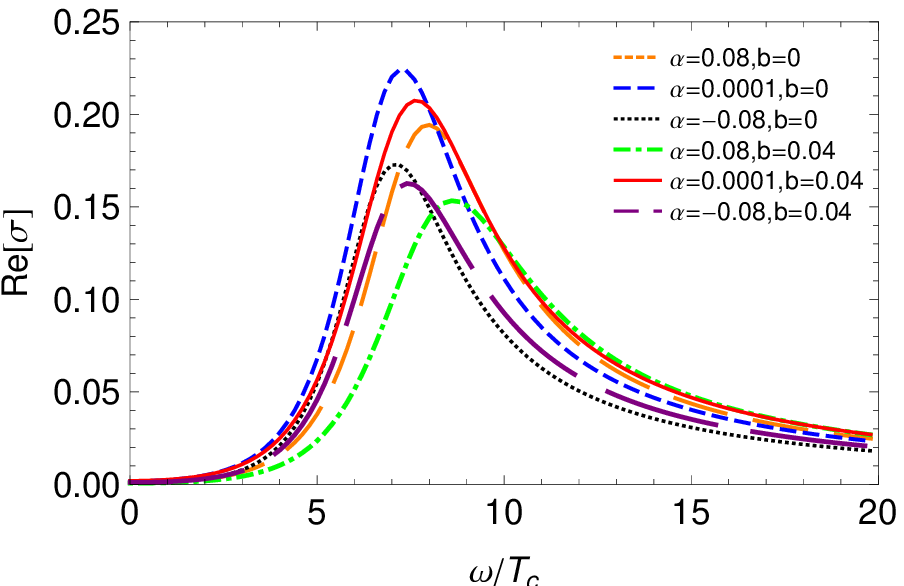}} \qquad %
\subfigure[~$\overline{m}^{2}=-3/4$]{\includegraphics[width=0.4\textwidth]{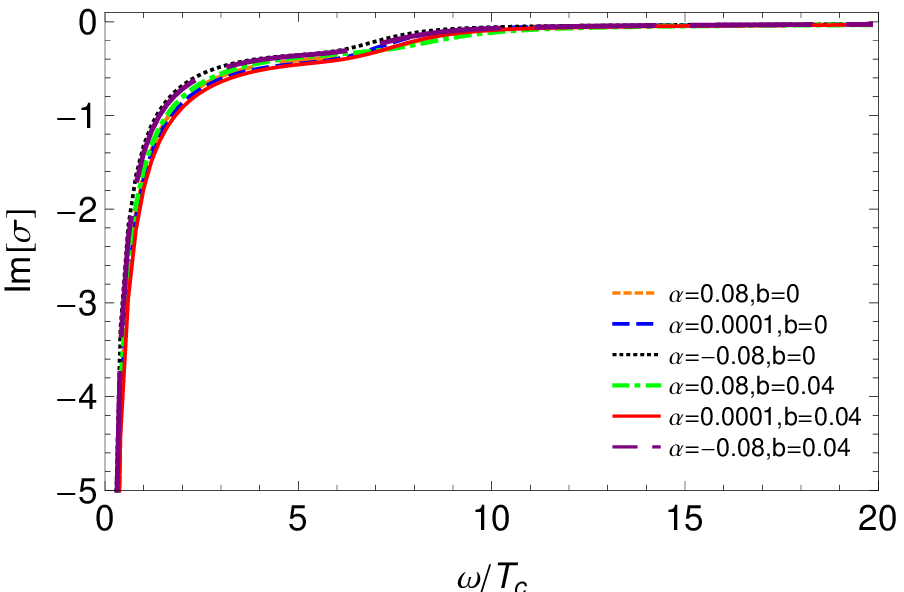}} \qquad %
\subfigure[~$\overline{m}^{2}=0$]{\includegraphics[width=0.4\textwidth]{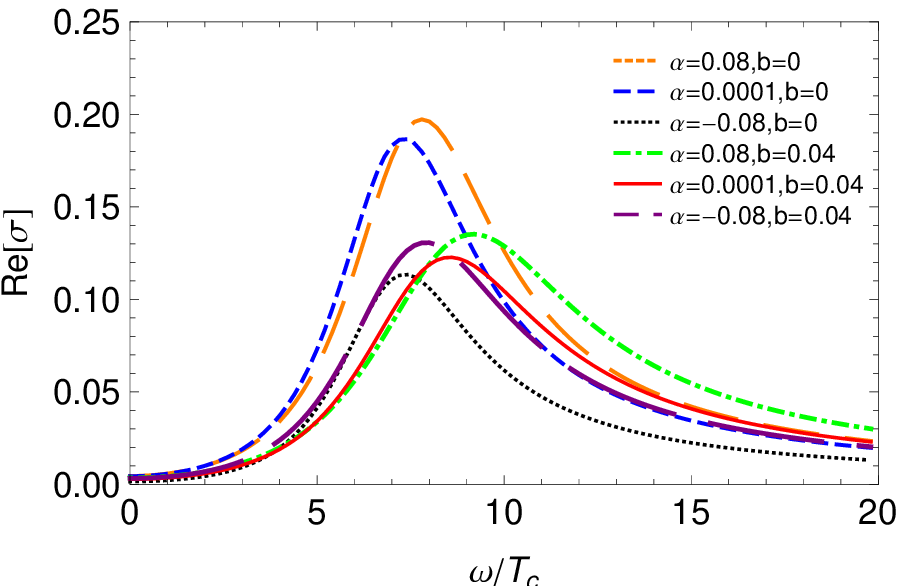}} \qquad %
\subfigure[~$\overline{m}^{2}=0$]{\includegraphics[width=0.4\textwidth]{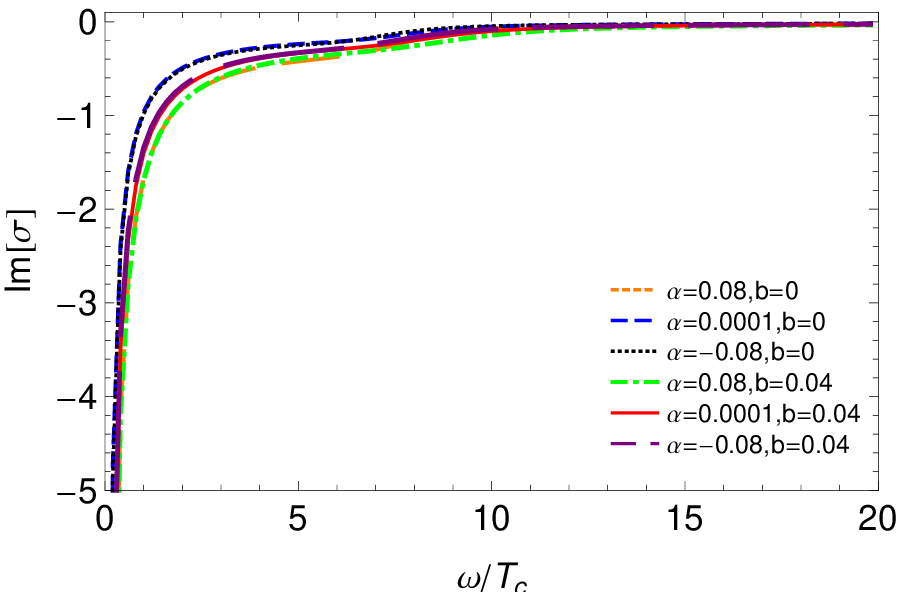}} \qquad %
\caption{The behavior of real and imaginary parts of conductivity
with $T/T_{c}=0.3$ in $D=3$ in $s$-wave case.}
\label{fig7}
\end{figure*}

Studying electrical conductivity is possible through applying an appropriate electromagnetic perturbation as $\delta A_{x} = A_{x}(r) e^{−i\omega t}$ on the black hole background which corresponds to the boundary electrical current in holographic setup\cite{H08}. Turning on this component yields to
\begin{equation}\label{as}
A_{x}''(r)+A_{x}'(r) \left[\frac{2 b \phi '(r) \phi ''(r)}{b \phi '(r)^2+1}+\frac{f'(r)}{f(r)}+\frac{(D-4)}{r}\right]+A_{x}(r) \left[\frac{\omega ^2}{f(r)^2}-\frac{2 \psi (r)^2}{f(r) \left(b \phi '(r)^2+1\right)}\right]=0,
\end{equation}
with asymptotic behavior as
\begin{equation}\label{eqasyms}
A_x''(r)+\frac{(D-2)}{r}A_x'(r)+\frac{\omega ^2 L_{\text{eff}}^4}{r^4} A_x(r)=0,
\end{equation}
which has the solution as

\begin{equation} \label{aysols}
A_{x} = \left\{
\begin{array}{lr}
A^{(0)}+A^{(1)} log\left(\dfrac{1}{r}\right)+\cdots, & D=3\\
\bigskip\\
A^{(0)}+\dfrac{A^{(1)}}{r}+\cdots, & D=4\\
\end{array} \right.
\end{equation}%
where $A^{(0)}$ and $A^{(1)}$ are constant parameters. Based on Ohm's law, electrical conductivity is obtained as
\begin{equation}\label{cons}
\sigma=-\frac{i A^{(1)}}{\omega A^{(0)}}
\end{equation}
which is in complete agreement with in Einstein gravity\cite{H08,mahyap11}. To analyze the behavior of conductivity as a function of frequency, we impose the ingoing wave boundary condition according to \cite{hscgb4}
\begin{equation}\label{ayexpand}
A_{y}(r)=f(r)^{-i \omega/ (4 \pi T)} \left[1+a(1-r)+b(1-r)+\cdots\right],
\end{equation}

in which $a$ and $b$ are constant parameters. Figures \ref{fig3}-\ref{fig6} show the trend of
real and imaginary parts of conductivity as a function of $\omega/T$ for different values of mass, nonlinearity and Gauss-Bonnet parameters in $D=3$ and $D=4$.
In general, the behavior of both parts of conductivity in $D=3$ is different from higher dimensions. At low frequency regime, we observe a delta function behavior in real part while imaginary part has a pole. At low temperatures, we face with appearance of gap in real part which occurs at about $\omega_{g}\approx8T_{c}$ based on figure \ref{fig7}. Increasing the effect of nonlinearity and Gauss-Bonnet parameters causes to deviate from this value. However, this isn't observed in $T/T_{c}=0.9$ case and in all temperatures of imaginary part. In high frequency limit, both parts approach to zero.

\begin{figure*}[t]
\centering
\subfigure[~$\overline{m}^{2}=-5/4$]{\includegraphics[width=0.4\textwidth]{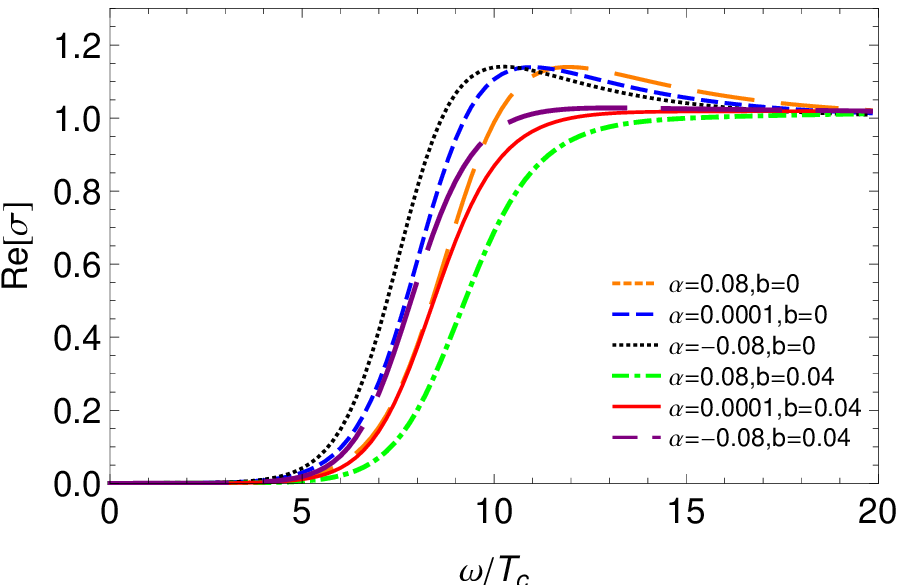}} \qquad %
\subfigure[~$\overline{m}^{2}=-5/4$]{\includegraphics[width=0.4\textwidth]{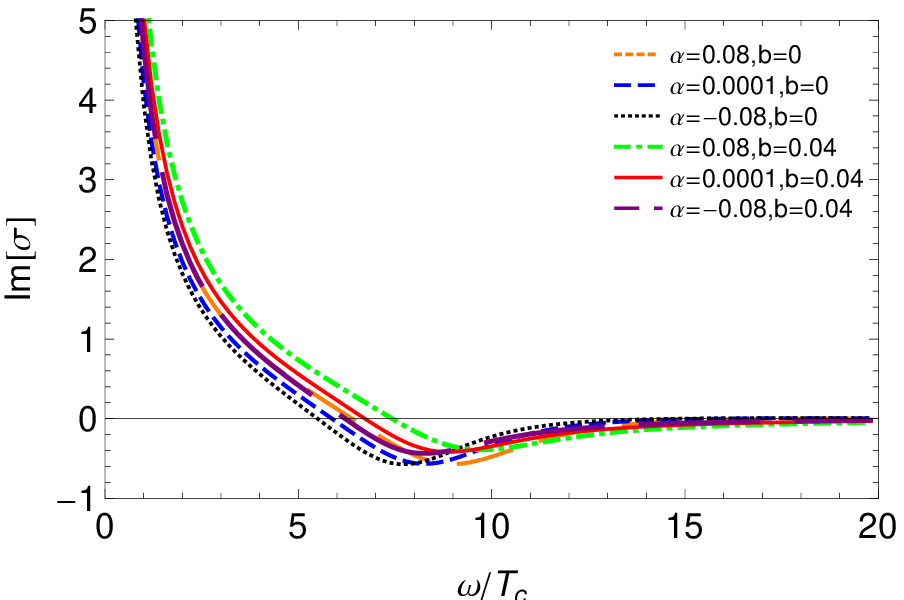}} \qquad %
\subfigure[~$\overline{m}^{2}=0$]{\includegraphics[width=0.4\textwidth]{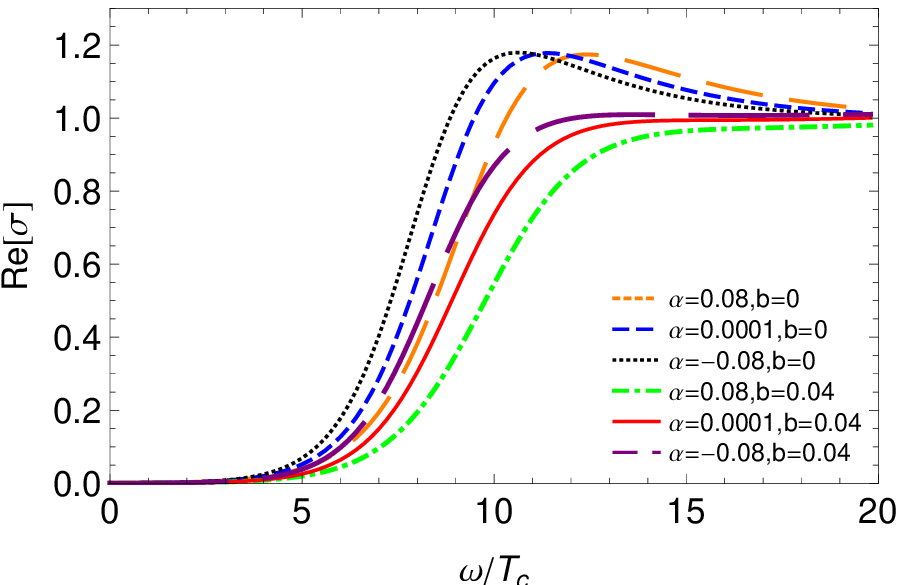}} \qquad %
\subfigure[~$\overline{m}^{2}=0$]{\includegraphics[width=0.4\textwidth]{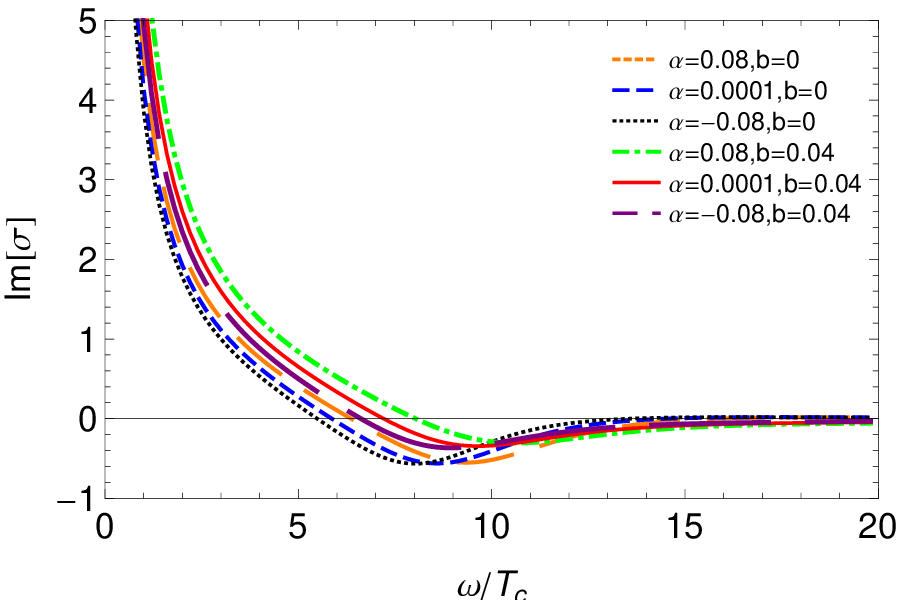}} \qquad %
\caption{The behavior of real and imaginary parts of conductivity
with $T/T_{c}=0.3$ in $D=4$ in $s$-wave case.}
\label{fig8}
\end{figure*}

The graphs in $D=4$ follow the same behavior as in the presence of Einstein gravity \cite{H08} by connecting the real and imaginary parts via Kramers-Kronig relation which means the occurrence of delta function and pole in real and imaginary parts of conductivity in $\omega \rightarrow 0$ limit, respectively.
At high frequency limit, both parts of conductivity tend to a constant value.
Furthermore, we observe that the superconducting gap in temperatures below the critical value becomes deeper and sharper by decreasing the temperature which leads to larger values of $\omega_{g}$ and indicates harder formation of fermion pairs and postpones the conductor/superconductor phase transition\cite{caipp}. Same trend is seen in previous section about condensation by diminishing the temperature.
In addition, in holographic setup $\omega_{g}\approx8T_{c}$ which differs from the BCS's value, $3.5$, due to strong coupling between the pairs which makes them good candidates to explore the behavior of high temperature superconductors. However, we face with deviation from $8$ to larger values for stronger effects of nonlinearity and Gauss-Bonnet terms.
To seek the effect of nonlinear electrodynamics and Gauss-Bonnet correction on the conductivity, we present the behavior of real and imaginary parts in fixed temperature $T=0.3T_{c}$ in figure \ref{fig8}. In general, the gap frequency is shifted in the presence of these two parameters.

\section{Holographic $p$-wave superconductor in Gauss-Bonnet gravity }\label{sectionp}
\section*{Condensation of a vector field} \label{sec5}

In order to study holographic $p$-wave superconductors with a vector field $\rho_{\mu}$ with mass $m$ and charge $q$, we present the action of matter field as
\begin{equation}\label{actmp}
S_{m} =\int d^{D}x\sqrt{-g} \left[
\mathcal{L}_{\mathcal{NL}}-\frac{1}{2}\rho_{\mu\nu}^{\dagger}
\rho^{\mu\nu}-m^{2} \rho_{\mu}^{\dagger} \rho^{\mu} + i q \gamma
\rho_{\mu} \rho_{\nu}^{\dagger} F^{\mu\nu}\right],
\end{equation}
in which $\mathcal{L}_{\mathcal{NL}}$, has the same definition as equation(\ref{actms}). After variation of action (\ref{actmp}) with respect to gauge and vector fields and substituting $A_{\nu}dx^{\nu}=\phi(r) dt $ and $\rho_{\nu}dx^{\nu}=\rho_{x}dx$ in the consequent equations, we get
\begin{table*}[t]
\label{tab3}
\begin{center}
\begin{tabular}{c|c|c|c|}
\cline{2-4}
& $b=0$ &$b=0.02$ & $b=0.04$ \\
\hline
\multicolumn{1}{|c|}{$\alpha=0.08$} & $0.0474 $ $\mu$& $0.0430 $ $\mu$ & $0.0396 $ $\mu$ \\
\hline
\multicolumn{1}{|c|}{$\alpha=0.0001$} & $0.0503 $ $\mu$ & $0.0461 $ $\mu$ & $ 0.0429$ $\mu$ \\
\hline
\multicolumn{1}{|c|}{$\alpha=-0.08$} & $0.0523 $ $\mu$ & $0.0488 $ $\mu$ & $ 0.0456$ $\mu$\\
\hline
\end{tabular}%
\caption{Numerical results for critical temperature $T_{c}$ with
$\overline{m}^{2}=1$ in $D=3$ in $p$-wave case for different values of nonlinearity
and Gauss-Bonnet parameters.}
\end{center}
\end{table*}

\begin{figure*}[t]
\centering
\subfigure[~$\overline{m}^{2}=1$]{\includegraphics[width=0.4\textwidth]{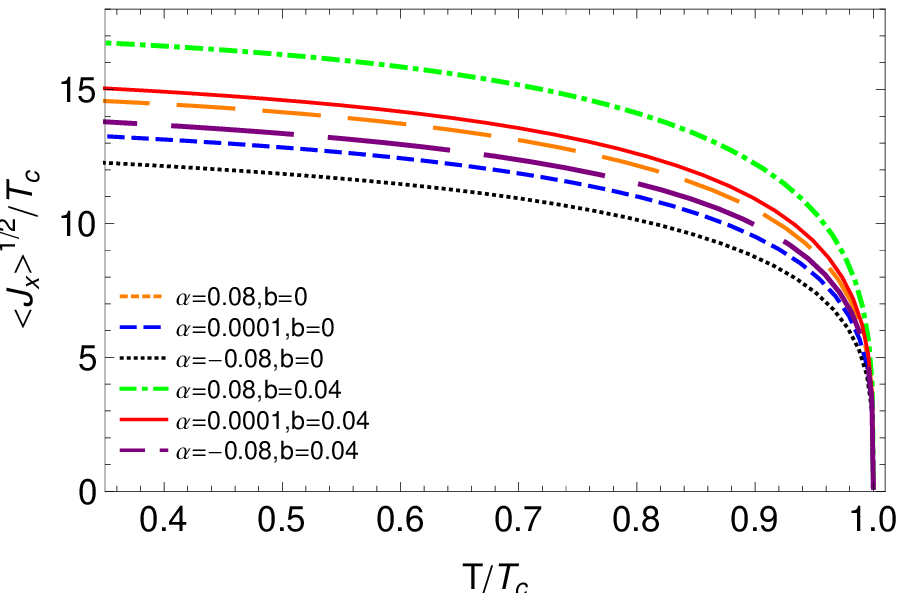}} \qquad %
\caption{The behavior of the condensation parameter as a function
of the temperature for different values of nonlinearity parameters
in $D=3$ in $p$-wave case.} \label{fig9}
\end{figure*}

\begin{figure*}[t]
\centering
\subfigure[~$\overline{m}^{2}=0$]{\includegraphics[width=0.4\textwidth]{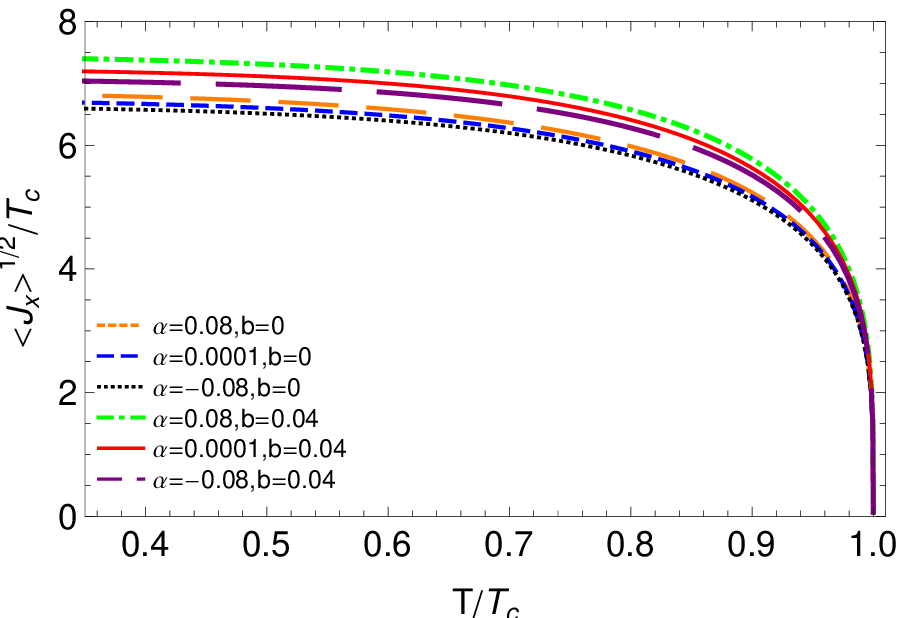}} \qquad %
\subfigure[~$\overline{m}^{2}=3/4$]{\includegraphics[width=0.4\textwidth]{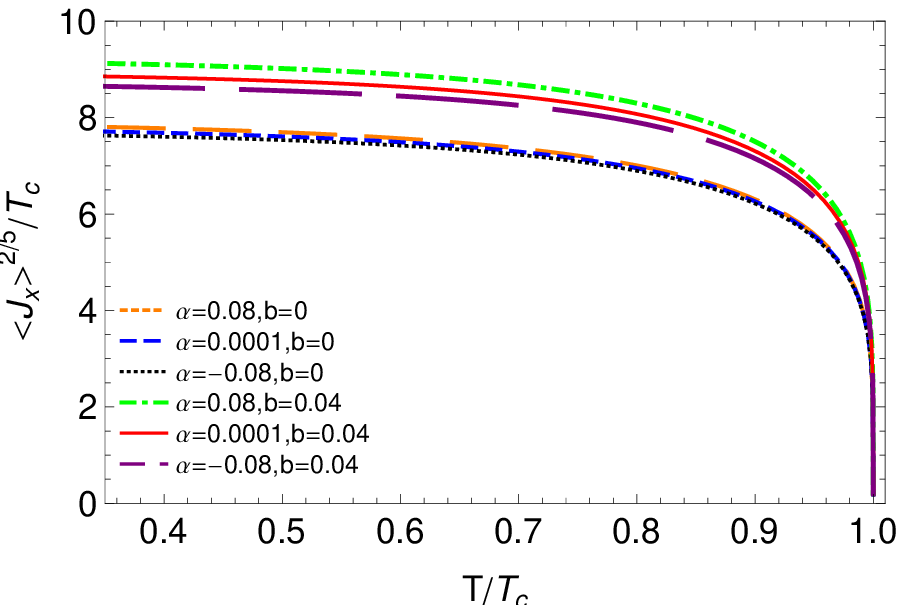}} \qquad %
\caption{The behavior of the condensation parameter as a function
of the temperature for different values of Gauss-Bonnet parameters
in $D=4$ in $p$-wave case.} \label{fig10}
\end{figure*}

\begin{table*}[t]
\label{tab4}
\begin{center}
\begin{tabular}{c|c|c|c|c|c|c|}
\cline{2-3}\cline{2-7}\cline{4-7}
& \multicolumn{2}{|c|}{$b=0$} & \multicolumn{2}{|c|}{$b=0.02$} &
\multicolumn{2}{|c|}{$b=0.04$} \\ \cline{2-3}\cline{2-7}\cline{4-7}
& $\overline{m}^{2}=0$ & $\overline{m}^{2}=3/4$ & $\overline{m}^{2}=0$ & $\overline{m}^{2}=3/4$ & $\overline{m}^{2}=0$ & $\overline{m}^{2}=3/4$
\\ \hline
\multicolumn{1}{|c|}{$\alpha=0.08$} & $ 0.1200$ $\sqrt{\rho}$ & $0.0980 $  $\sqrt{\rho}$ & $0.1138 $  $\sqrt{\rho}$ & $ 0.0881$  $\sqrt{\rho}$ & $0.1094 $  $\sqrt{\rho}$ & $0.0821 $  $\sqrt{\rho}$ \\ \hline
\multicolumn{1}{|c|}{$\alpha=0.0001$} & $0.1250 $  $\sqrt{\rho}$ & $0.1021 $  $\sqrt{\rho}$ & $0.1194 $  $\sqrt{\rho}$ & $0.0931 $  $\sqrt{\rho}$ & $0.1153 $  $\sqrt{\rho}$ & $ 0.0873$  $\sqrt{\rho}$ \\ \hline
\multicolumn{1}{|c|}{$\alpha=-0.08$} & $0.1291 $  $\sqrt{\rho}$  & $0.1055 $  $\sqrt{\rho}$ & $0.1241 $  $\sqrt{\rho}$ & $%
0.0971$  $\sqrt{\rho}$ & $0.1202 $ $\sqrt{\rho}$ & $0.0916 $  $\sqrt{\rho}$ \\ \hline
\end{tabular}%
\caption{Numerical results for critical temperature $T_{c}$ in $D=4$ in $p$-wave case for different values of mass, nonlinearity and Gauss-Bonnet parameters}
\end{center}
\end{table*}

\begin{equation}\label{phip}
\phi ''(r)+\frac{\phi '(r) (D-2)}{r} \left[\frac{b \phi '(r)^2+1}{3 b \phi '(r)^2+1}\right]-\frac{2 q^2 \rho_{x} (r)^2 \phi (r)}{r^2 f(r)}\left[\frac{1}{ 3 b \phi '(r)^2+1}\right]=0,
\end{equation}
\begin{equation}\label{rho}
\rho_{x} ''(r)+\rho_{x} '(r)\left[\frac{f'(r)}{f(r)}+\frac{(D-4)}{r}\right] +\rho_{x} (r) \left[\frac{q^2 \phi (r)^2}{f(r)^2}-\frac{m^2}{f(r)}\right]=0,
\end{equation}
in the Maxwell limit, $b\rightarrow0$, field equations turn to corresponding equations in \cite{mahyagaussp}. The asymptotic behavior of equation(\ref{phip}) is same as equation(\ref{asymphis}). However, we now define the form of equation(\ref{rho}) in $r\rightarrow \infty$ limit as
\begin{equation}\label{eqasymrho}
\rho_{x}(r)=\dfrac{\rho_{x+}(r)}{r^{\Delta_{+}}}+\dfrac{\rho_{x-}(r)}{r^{\Delta_{-}}}, \  \   \  \  \   \Delta _\pm=\frac{1}{2} \left[(D-3)\pm\sqrt{(D-3)^2+4 m^2 L_{\text{eff}}^{2}}\right],
\end{equation}
whith the Breitenlohner-Freedman (BF) bound as
\begin{equation}\label{bfp}
\overline{m}^{2}\geqslant - \frac{(D-3)^{2}}{4}, \ \ \ \ \overline{m}^{2}=m^{2} L_{\text{eff}}^{2}.
\end{equation}
Based on the AdS/CFT duality $\rho_{x-}$ and $\rho_{x+}$ are interpreted as the source and the expectation value $\langle J_{x}\rangle$ which is considered as order parameter in the boundary theory. By following the same approach as in section \ref{sec3}, we summarize the results of numerical solutions affected by different values of mass, nonlinearity and Gauss-Bonnet parameters in tables III and IV. We observe that the critical temperature goes down by going up each of these terms. Furthermore, in $D=4$ with $\alpha=0.0001$, the results of critical temperature in \cite{mahyagaussp} are obtained which prove the point that in small regime of Gauss-Bonnet coefficient the outcomes of Einstein case are recovered.
Due to limitation of numerical solution, we can do our calculation only with $\overline{m}^{2}=1$ for holographic $p$-wave superconductor in $D=3$.
The effect of mass, nonlinearity and Gauss-Bonnet parameters are illustrated in figures \ref{fig9}-\ref{fig10} by showing the behavior of $\langle J_{x} \rangle^{1/(1+\Delta_{+})}$ as a function of temperature . In all graphs increasing each of these three parameters hinders conductor/superconductor phase transition. The trend of figures are the same as BCS theory with a square root behavior which is the sign of the second order phase transition. According to these graphs condensation values increase because of going up these terms which means it is harder to have a superconductor in the presence of strong nonlinear and Gauss-Bonnet terms for massive vector fields.

\section*{Conductivity}\label{sec6}
\begin{figure*}[t]
\centering
\subfigure[~$\alpha=0.08$]{\includegraphics[width=0.4\textwidth]{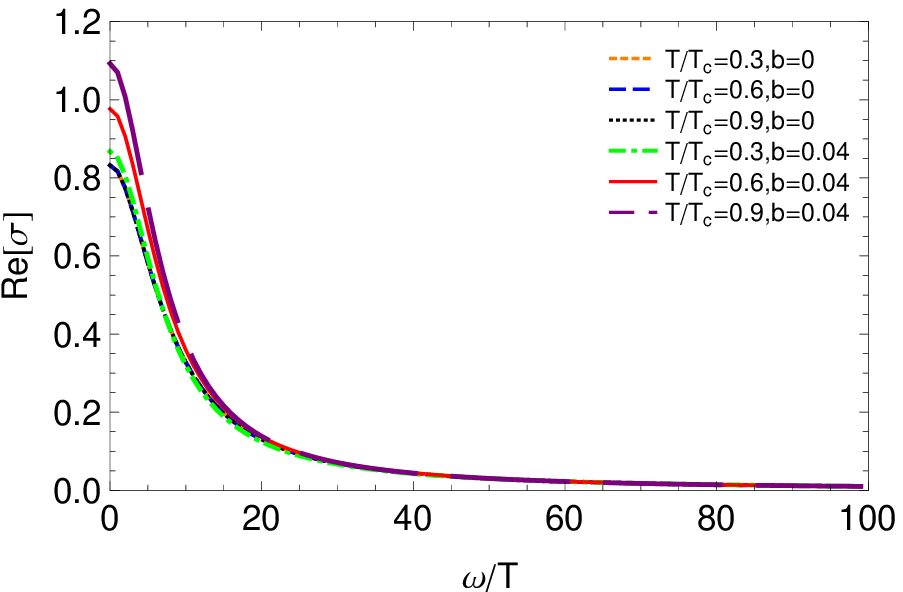}} \qquad %
\subfigure[~$\alpha=-0.08$]{\includegraphics[width=0.4\textwidth]{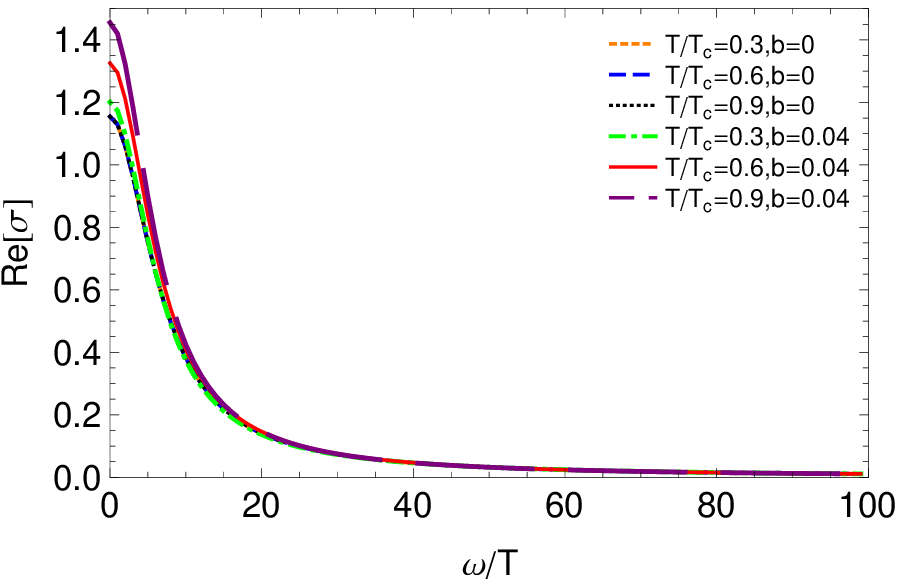}} \qquad %
\subfigure[~$\alpha=0.08$]{\includegraphics[width=0.4\textwidth]{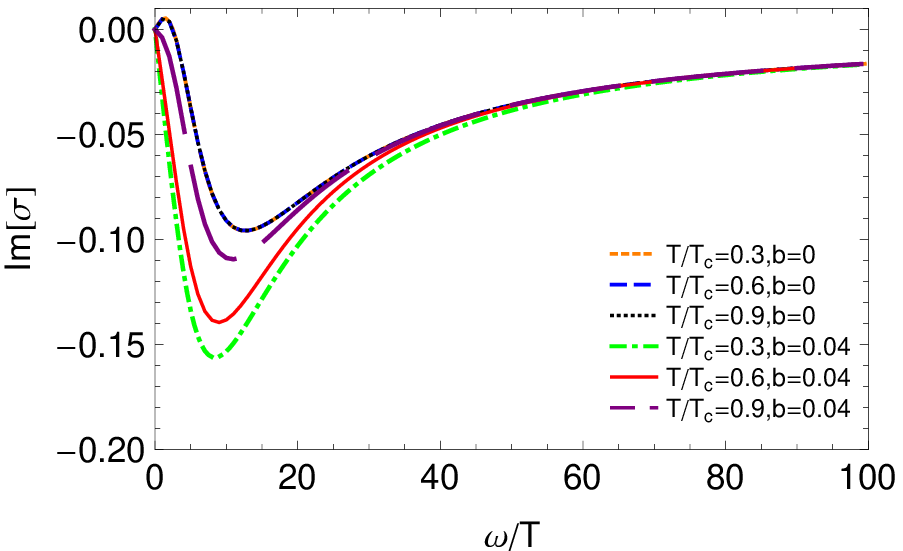}} \qquad %
\subfigure[~$\alpha=-0.08$]{\includegraphics[width=0.4\textwidth]{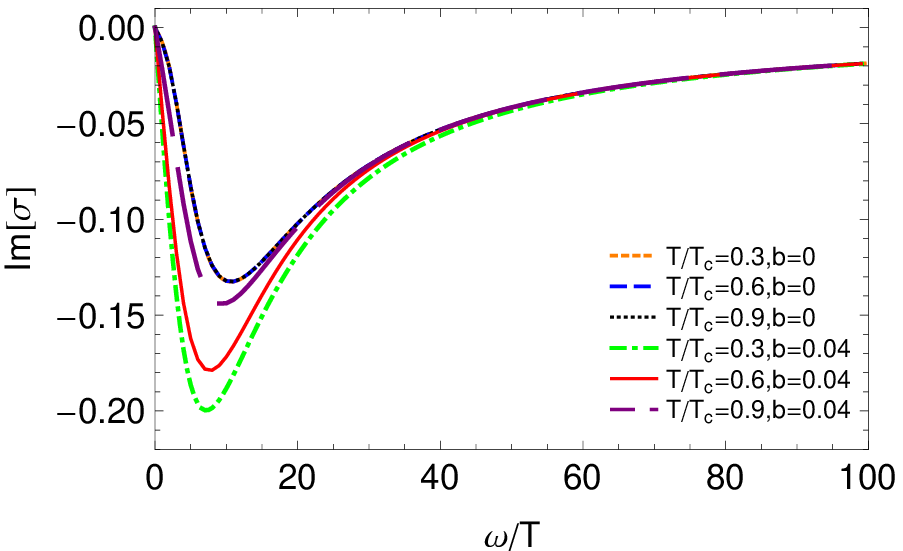}} \qquad %
\caption{The behavior of real and imaginary parts of conductivity with
$\overline{m}^{2}=1$ in $D=3$ in $p$-wave case. } \label{fig11}
\end{figure*}

\begin{figure*}[t]
\centering
\subfigure[~ $\alpha=0.08$]{\includegraphics[width=0.4\textwidth]{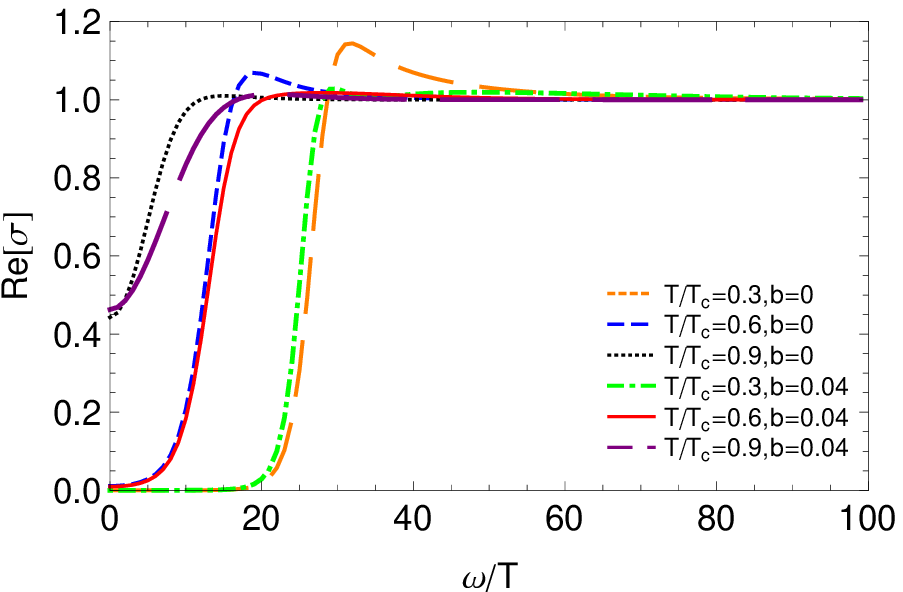}} \qquad %
\subfigure[~ $\alpha=-0.08$]{\includegraphics[width=0.4\textwidth]{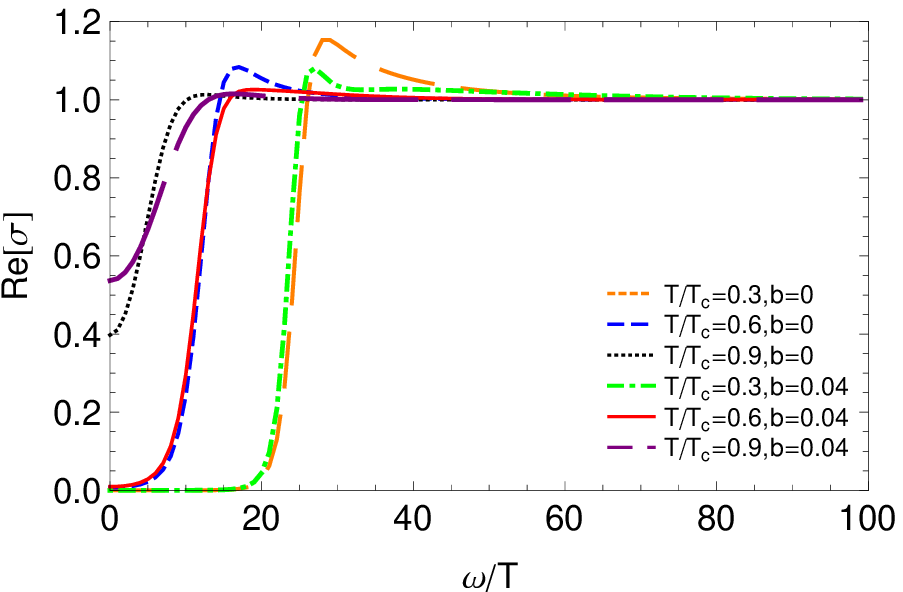}} \qquad %
\subfigure[~ $\alpha=0.08$]{\includegraphics[width=0.4\textwidth]{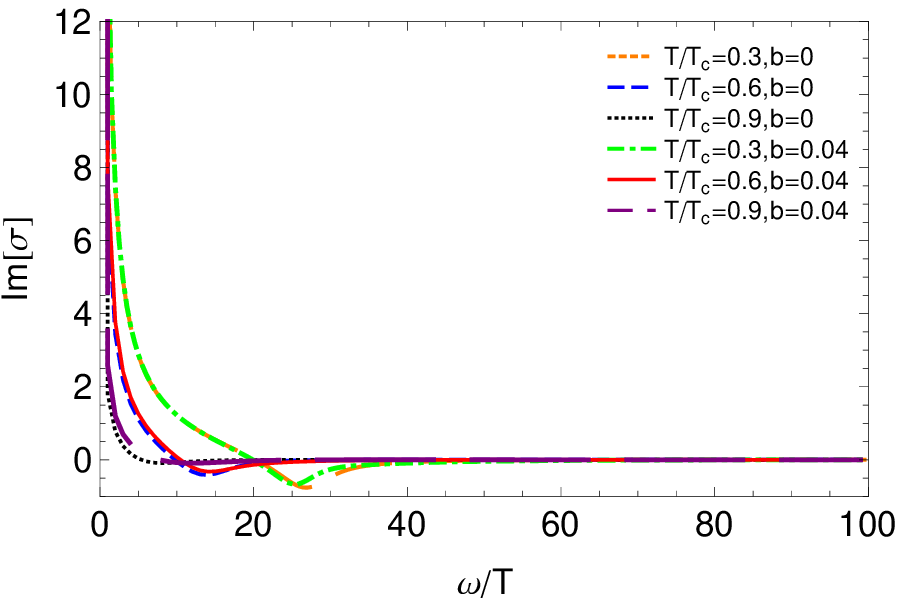}} \qquad %
\subfigure[~$\alpha=-0.08$]{\includegraphics[width=0.4\textwidth]{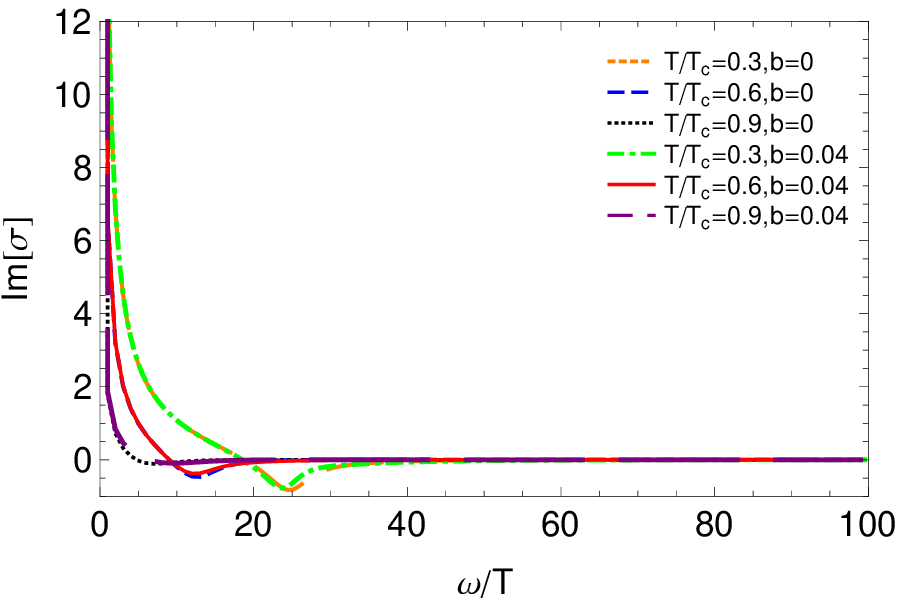}} \qquad %
\caption{The behavior of real and imaginary parts of conductivity with
$\overline{m}^{2}=0$ in $D=4$ in $p$-wave case.} \label{fig12}
\end{figure*}

\begin{figure*}[t]
\centering
\subfigure[~$\alpha=0.08$]{\includegraphics[width=0.4\textwidth]{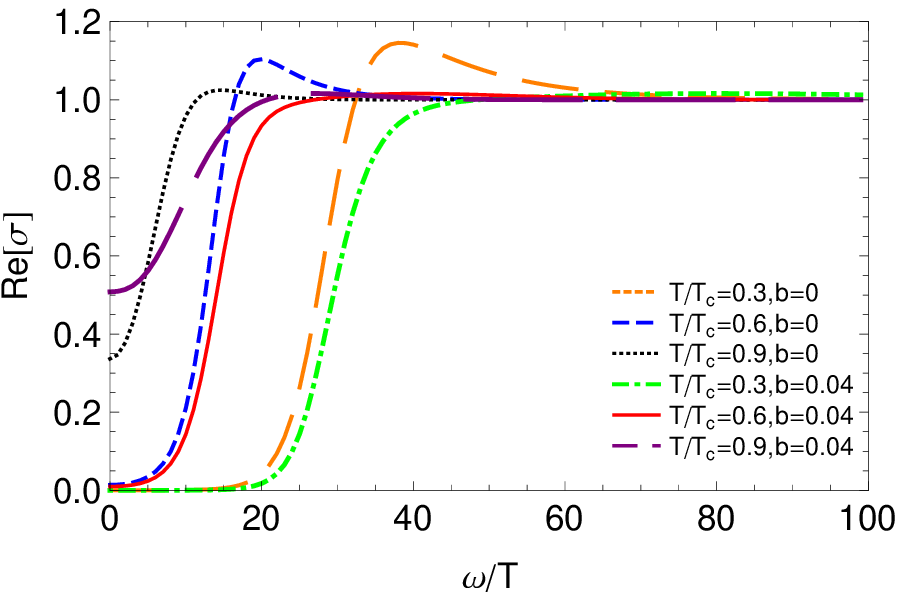}} \qquad %
\subfigure[~$\alpha=-0.08$]{\includegraphics[width=0.4\textwidth]{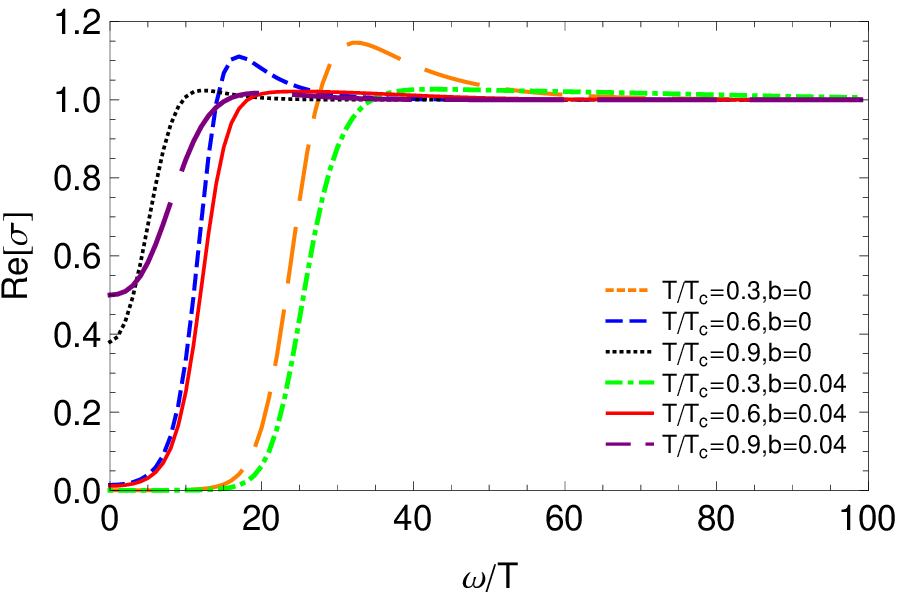}} \qquad %
\subfigure[~$\alpha=0.08$]{\includegraphics[width=0.4\textwidth]{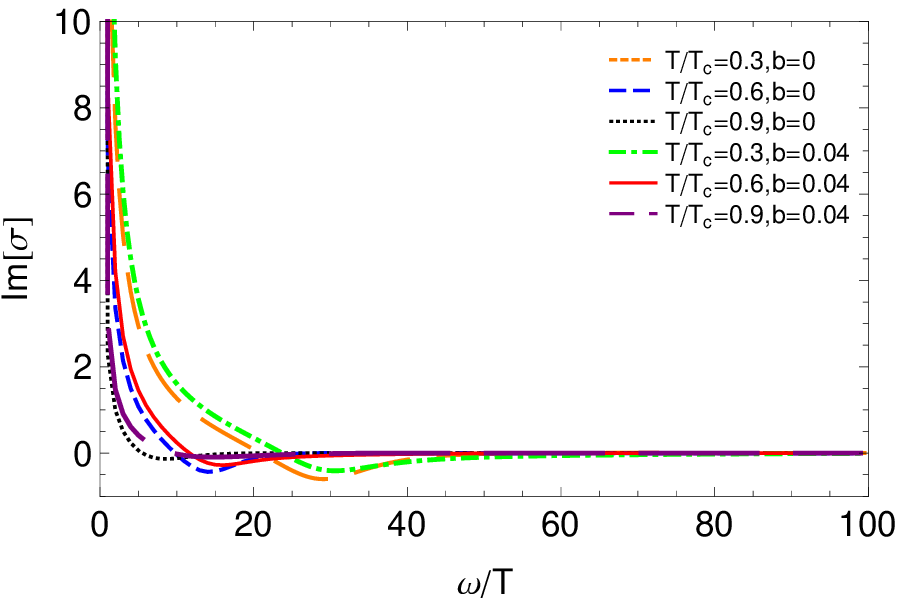}} \qquad %
\subfigure[~$\alpha=-0.08$]{\includegraphics[width=0.4\textwidth]{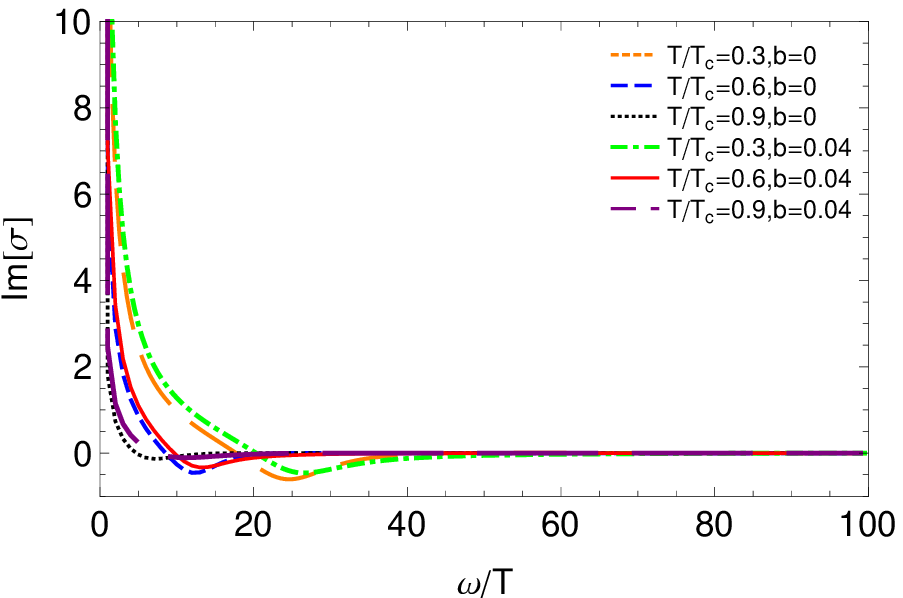}} \qquad %
\caption{The behavior of real and imaginary parts of conductivity with
$\overline{m}^{2}=3/4$ in $D=4$ in $p$-wave case.} \label{fig13}
\end{figure*}

\begin{figure*}[t]
\centering
\subfigure[~$\overline{m}^{2}=1$]{\includegraphics[width=0.4\textwidth]{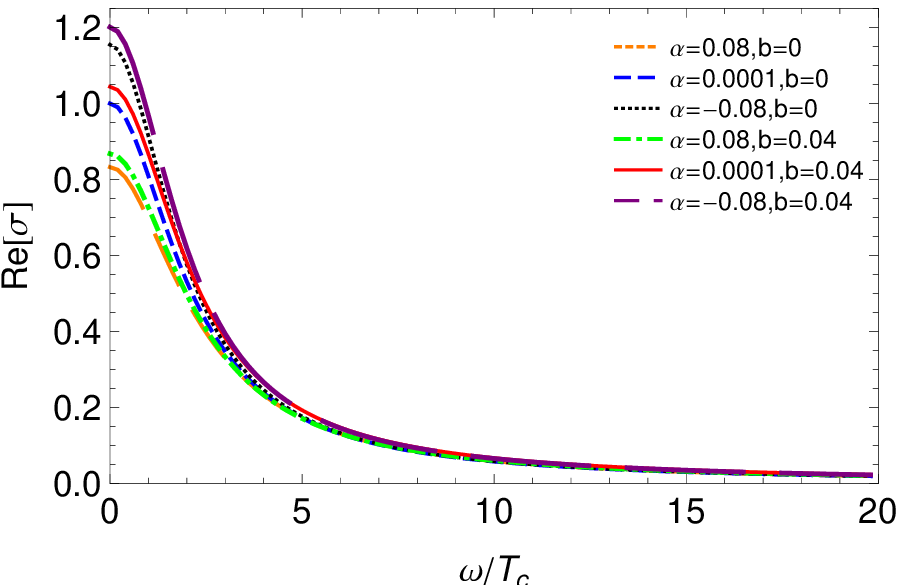}} \qquad %
\subfigure[~$\overline{m}^{2}=1$]{\includegraphics[width=0.4\textwidth]{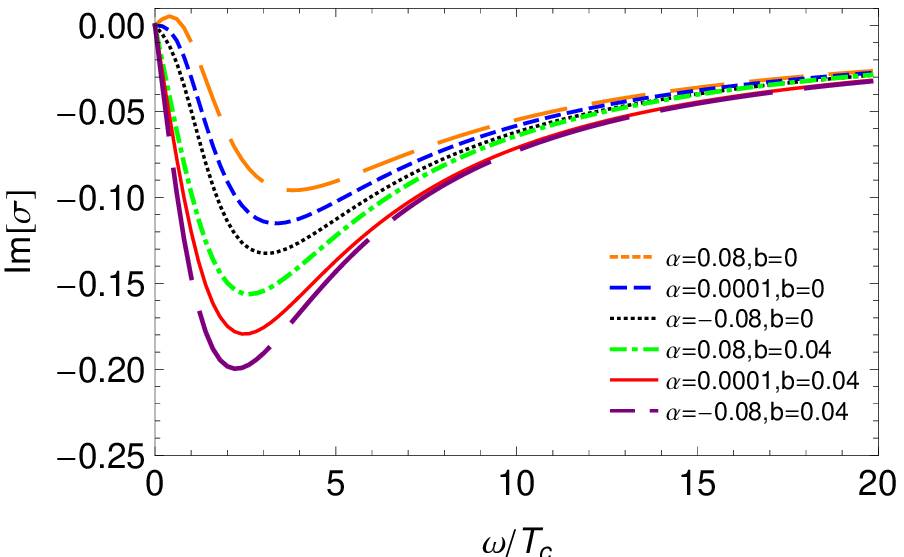}} \qquad %

\caption{The behavior of real and imaginary parts of conductivity
with $T/T_{c}=0.3$ in $D=3$ in $p$-wave case.}
\label{fig14}
\end{figure*}

\begin{figure*}[t]
\centering
\subfigure[~$\overline{m}^{2}=0$]{\includegraphics[width=0.4\textwidth]{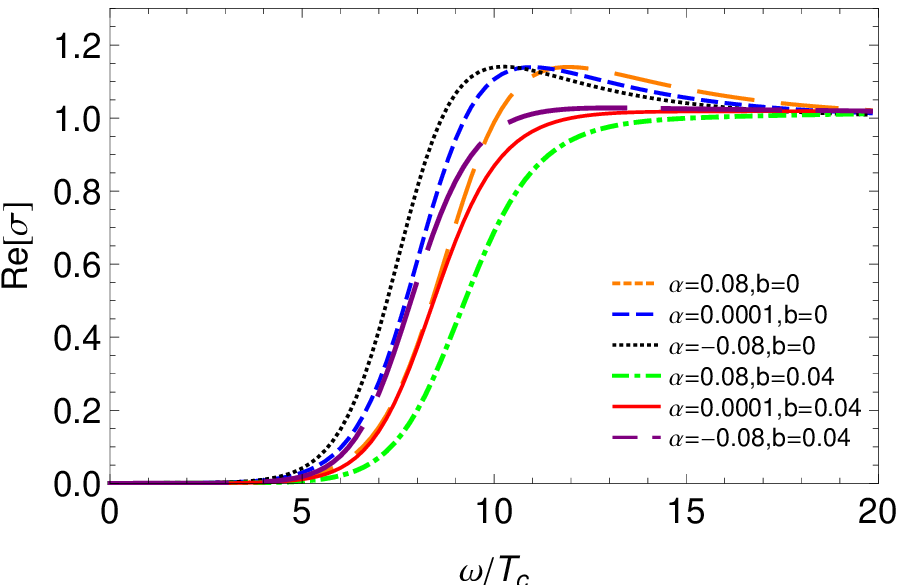}} \qquad %
\subfigure[~$\overline{m}^{2}=0$]{\includegraphics[width=0.4\textwidth]{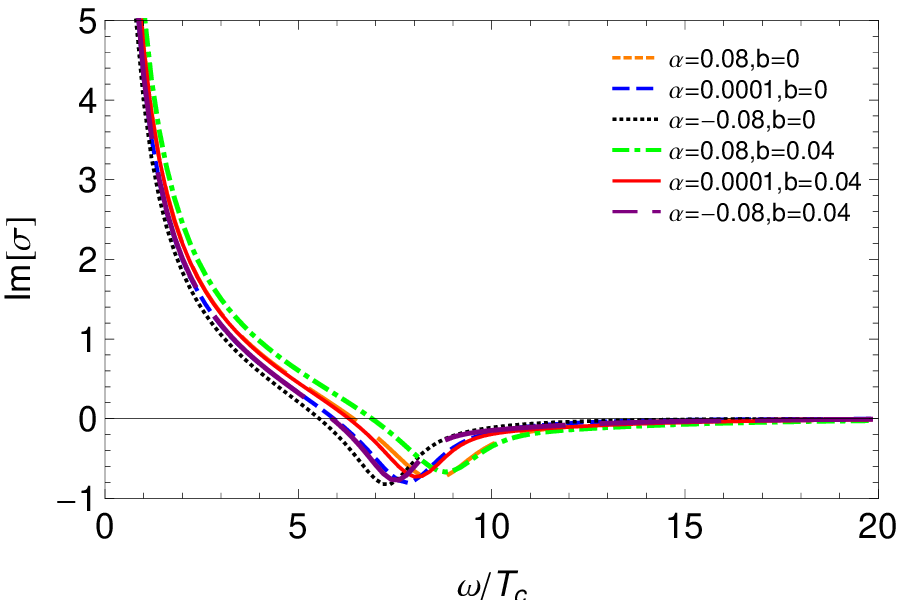}} \qquad %
\subfigure[~$\overline{m}^{2}=3/4$]{\includegraphics[width=0.4\textwidth]{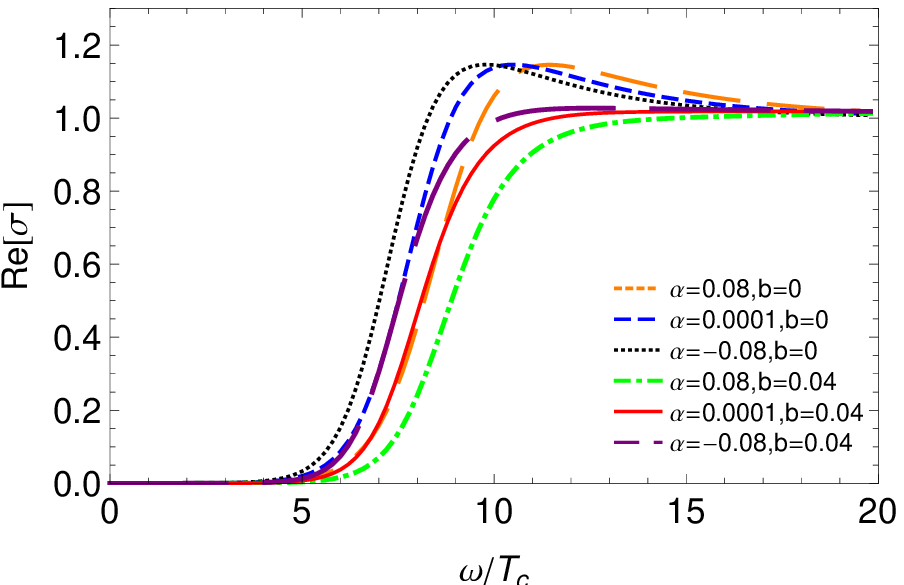}} \qquad %
\subfigure[~$\overline{m}^{2}=3/4$]{\includegraphics[width=0.4\textwidth]{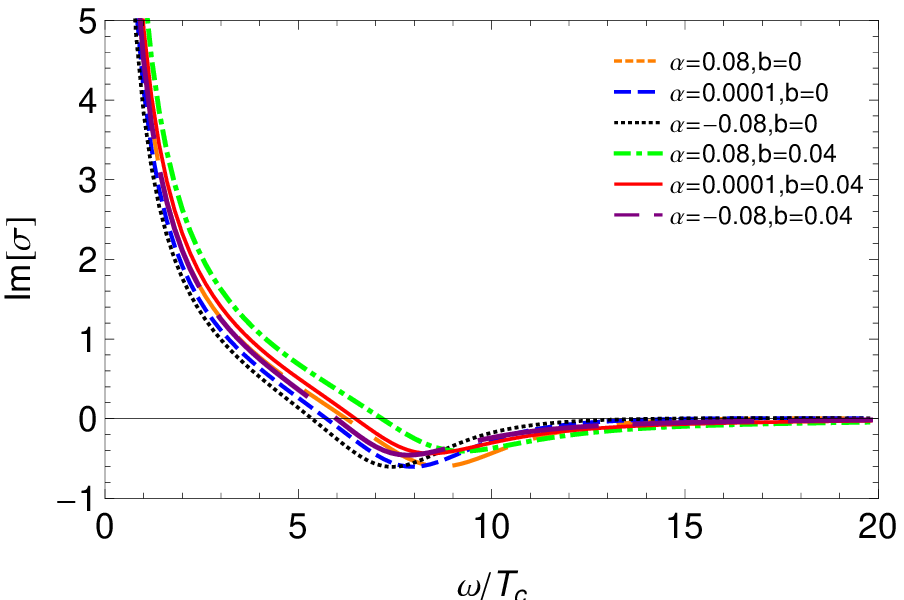}} \qquad %
\caption{The behavior of real and imaginary parts of conductivity
for $T/T_{c}=0.3$ in $D=4$ in $p$-wave case.}
\label{fig15}
\end{figure*}

To explore the conductivity of holographic $p$-wave superconductors, we apply the same perturbation as in section \ref{sec3} in $D=3$ but for $D=4$ in order to prevent complexity we turn on the $y$-component of gauge field as $\delta A_{y}= A_{y} e^{-i \omega t}$ which yields to
\begin{equation}\label{ap}
A_{x(y)}''(r)+A_{x(y)}'(r) \left[\frac{2 b \phi '(r) \phi ''(r)}{b \phi '(r)^2+1}+\frac{f'(r)}{f(r)}+\frac{(D-4)}{r}\right]+A_{x(y)}(r) \left[\frac{\omega ^2}{f(r)^2}-\frac{2 (D-3) \rho_{x}(r)^2}{ r^2 f(r) \left(b \phi '(r)^2+1\right)}\right]=0,
\end{equation}
in which indices $x$ and $y$ belong to $D=3$ and $D=4$ respectively. The asymptotic behavior of the above equation has the following form
\begin{equation}\label{eqasymayp}
A_{x(y)}''(r)+\frac{(D-2)}{r}A_{x(y)}'(r)+\frac{\omega ^2 L_{\text{eff}}^4}{r^4} A_{x(y)}(r)=0,
\end{equation}
which has the solution as
\begin{equation} \label{aysolp}
\left\{
\begin{array}{lr}
A_{x} =A^{(0)}+A^{(1)} log\left(\dfrac{1}{r}\right)+\cdots, & D=3\\
\bigskip\\
A_{y} =A^{(0)}+\dfrac{A^{(1)}}{r}+\cdots, & D=4\\
\end{array} \right.
\end{equation}%
with $A_{0}$ and $A_{1}$ as constant parameters. So, the electrical conductivity has the same definition as equation(\ref{cons}).
 As said before the behavior of conductivity in $D=3$ is so different from higher dimensions which is shown in figure \ref{fig11}. Even the behavior of real and imaginary parts differs in $s$-wave and $p$-wave cases. In low frequency limit, we observe a delta function behavior in real part while the imaginary part contrary to other cases tends to zero. In addition, we face with absence of gap in real part while imaginary section shows a minimum which becomes deeper by going up the Gauss-Bonnet and nonlinearity terms. So, these two parts are not connected to each other through Kramers-Kronig equation. In high frequency region, both parts tend to a constant value.
In $D=4$, the behavior of conductivity is exactly the same as in $s$-wave model which shows that turning on the $y$-component of $A_{\mu}$ in $p$-wave case has the same effect as $x$ term of $A_{\mu}$ in $s$-wave model. By paying attention to figures \ref{fig12}-\ref{fig13} which show the behavior of real and imaginary parts of conductivity as a function of frequency in $D=4$, we realize that the effects of mass, Gauss-Bonnet and nonlinearity parameters play exactly the same role in holographic $s$- and $p$-wave superconductors which explained in section \ref{sec3}. Moreover, The dependence of conductivity to nonlinearity and Gauss-Bonnet parameters at $T/T_{c}=0.3$ is shown in figures \ref{fig14}-\ref{fig15}.

\section{summary and conclusion}\label{section5}
In this work by applying AdS/CFT duality, we study holographic $s$- and $p$-wave superconductors by considering the higher order corrections in gravity and gauge field sides in $D=3$ and $D=4$ by considering the action which is introduced in \cite{RBM1,RBM2}. Firstly by using shooting method, we analyze the equations of motion numerically and find the relation between critical temperature $T_{c}$ and chemical potential $\mu$ or charge density $\rho$ for different values of mass, nonlinear and Gauss-Bonnet terms numerically. Our results show that in all cases increasing these terms diminishes the critical temperature which makes the conductor/superconductor phase transition harder to form. In addition, we plot the behavior of condensation as a function of temperature to illustrate the effect of different values of mass, nonlinearity and Gauss-Bonnet parameters. We face with larger values of condensation due to increasing each of these three parameters which is the sign of harder superconductor formation and can be considered as a confirmation of the results of critical temperature. Secondly, we calculate the electrical conductivity in holographic setup by turning on an appropriate perturbation on black hole background which corresponds to electrical current in CFT part. We present the behavior of real and imaginary parts of conductivity affected by different values of mass, nonlinearity and Gauss-Bonnet parameters. Based on our results, the behavior of both parts of conductivity in holographic $s$- and $p$-wave superconductors in $D=4$ is exactly the same and follows the same behavior as higher dimensions in \cite{Afsoon,mahyagaussp}. In $\omega\rightarrow 0$ limit, real and imaginary parts are connected to each other via Kramers-Kronig relationship by appearance of delta function and divergence behavior respectively. One of superconductivity phase characteristic is infinite DC conductivity. Moreover in high frequency regime, both parts tend to a constant value. Based on the BCS prediction $\omega_{g}\approx3.5T_{c}$ while we obtain larger value around $8$ which deviates from this value in the presence of nonlinear electrodynamics and Gauss-bonnet correction for massive fields. Furthermore, the gap energy and minimum in real and imaginary parts shift toward larger frequencies by diminishing temperature and increasing the effect of nonlinear and Gauss-Bonnet terms. However, the behavior of conductivity in $D=3$ is so different from other dimensions. Even the graphs in $s$-wave case are far different from those in $p$-wave model. In holographic $s$-wave model, the behavior of both parts in low and high frequency regimes are alike those in $D=4$. However, at low temperatures we observe the occurrence of a gap in real part at about $\omega_{g}\approx8T_{c}$ while the imaginary part follows a gentle trend without a minimum. Conductivity of holographic $p$-wave superconductor in $D=3$ behave far different from the previous cases. For example, in $\omega\rightarrow 0$ region, both parts have a delta function behavior and imaginary part doesn't tend to infinity. Now, we see a minimum in imaginary part which becomes deeper for stronger effect of nonlinearity and Gauss-bonnet terms while the gap energy doesn't appear in real section. Both parts of conductivity at high frequency regime tend to a constant value which is similar to other cases. Maybe, considering the effect of backreaction can give more information about conductivity in $D=3$.
\section*{Data Availability Statement}
Data sharing not applicable to this article as no datasets were generated or analysed during the current study.


\begin{thebibliography}{Horowitz et al.(2008)}


\bibitem[Maldacena.(1998)]{Maldacena} J. M. Maldacena, Adv. Theor. Math.
Phys. \textbf{2}, 231 (1998) [arXiv:hep-th/9711200v3].



\bibitem{H08} S. A. Hartnoll, C. P. Herzog and G. T. Horowitz, Phys. Rev.
Lett. \textbf{101}, 031601 (2008) [arXiv:0803.3295 [hep-th]].



\bibitem[Gubser et al.(1998)]{G98} S. S. Gubser, I. R. Klebanov and A. M.
Polyakov, Phys. Lett. B \textbf{428}, 105 (1998) [arXiv:hep-th/9802109].






\bibitem[Witten et al.(1998)]{W98} E. Witten, Adv. Theor. Math. Phys.
\textbf{2}, 253 (1998) [arXiv:hep-th/9802150].






\bibitem[Horowitz et al.(2008)]{HR08} G. T. Horowitz and M. M. Roberts,
Phys. Rev. D \textbf{78}, 126008 (2008).



\bibitem[Ren.(2010)]{R10} J. Ren, JHEP. \textbf{1011}, 055 (2010)
[arXiv:1008.3904[hep-th]].



\bibitem[Horowitz(2011)]{H11} G. T. Horowitz, Lect. Notes Phys. \textbf{828}%
, 313 (2011) [arXiv:1002.1722[hep-th]].






\bibitem[Hartnoll(2009)]{H09} S. A. Hartnoll, Class. Quantum Grav. \textbf{26%
}, 224002 (2009) [arXiv:0903.3246[hep-th]].






\bibitem{25} M. Born and L. Infeld, Proc. R. Soc. A \textbf{144}, 425 (1934).
\bibitem{hendi} S.H. Hendi and A. Sheykhi, Phys. Rev. D, {\bf 88}, 044044 . [arXiv:1405.6998[gr-qc]].
\bibitem{log} H. H. Soleng, Phys. Rev. D {\bf 52}, 6178 (1995), [arXiv:hep-th/9509033].



\bibitem[Salahi et al.(2016)]{SSh16} H. R. Salahi, A. Sheykhi, A. Montakhab,
Eur. Phys. J. C \textbf{76}, 575 (2016) [arXiv:1608.05025[gr-qc]].
\bibitem[Herzog.(2009)]{Hg09} C. P. Herzog, J. Phys. A \textbf{42}, 343001
(2009) [arXiv:0904.1975[hep-th]].









\bibitem[Gubser.(2009)]{Gu09} S. S. Gubser, C. P. Herzog, S. S. Pufu and T.
Tesileanu, Phys. Rev. Lett. \textbf{103}, 141601 (2009) [arXiv:0907.3510[hep-th]].



\bibitem[HHH.(2008)]{HHH08} S. A. Hartnoll, C. P. Herzog and G. T. Horowitz,
JHEP \textbf{0812}, 015 (2008) [arXiv:0810.1563[hep-th]].



\bibitem[Jing ,Chena(210)]{JCH10} J. Jing, S. Chen, Phys. Lett. B \textbf{686%
}, 68 (2010) [arXiv:1001.4227[gr-qc]].



\bibitem[cai(2015)]{cai15} R. G. Cai, L. Li, Li-Fang Li, Run-Qiu Yang, Sci
China Phys. Mech. Astron. \textbf{58}, 060401 (2015) [arXiv:1502.00437[hep-th]].



\bibitem[Ge(2010)]{Ge10} X. H. Ge, B. Wang, S. F. Wu, and G. H. Yang, JHEP
\textbf{1008}, 108 (2010) [arXiv:1002.4901[hep-th]].



\bibitem[Ge(2012)]{Ge12} X. H. Ge, S. F. Tu, B. Wang, JHEP \textbf{09}, 088
(2012) [arXiv:1209.4272[hep-th]].



\bibitem[Kuang(2013)]{Kuang13} X. M. Kuang, E. Papantonopoulos, G. Siopsis,
B. Wang, Phys. Rev. D \textbf{88}, 086008 (2013) [arXiv:1303.2575[hep-th]].



\bibitem[Pan(2011)]{Pan11} Q. Pan, J. Jing, B. Wang, JHEP \textbf{11}, 088
(2011) [arXiv:1105.6153 [gr-qc]].






\bibitem[CAI(11)]{CAI11} R. G. Cai, H. F Li, H.Q. Zhang, Phys. Rev. D
\textbf{83}, 126007 (2011).



\bibitem[cai(10)]{cai10} R. G. Cai, Z.Y. Nie, H.Q. Zhang, Phys. Rev. D
\textbf{82}, 066007 (2010).






\bibitem[yao(2013)]{yao13} W. Yao, J. Jing, JHEP \textbf{1305}, 101 (2013)
[arXiv:1306.0064[gr-qc]].
\bibitem{Gan1} S. Gangopadhyay, D. Roychowdhury, JHEP \textbf{05}, 002
(2012) [arXiv:1201.6520[hep-th]]
\bibitem{Doa} A. Sheykhi, D. Hashemi Asl, A. Dehyadegari, Phys. Lett. B
\textbf{781}, 139 (2018) [arXiv:1803.05724[hep-th]].



\bibitem{Afsoon} A. Sheykhi, A. Ghazanfari, A. Dehyadegari, Eur. Phys. J. C
\textbf{78}, 159 (2018) [arXiv:1712.04331[hep-th]].



\bibitem{n4} Z. Zhao, Q. Pan, S. Chen and J. Jing, Nucl. Phys. B \textbf{871}%
, 98 (2013) [arXiv:1212.6693[hep-th]].



\bibitem{n6} Y. Liu, Y. Gong and B. Wang, JHEP \textbf{1602}, 116 (2016)
[arXiv:1505.03603[hep-ph]].









\bibitem[Sheykhi et al.(2016)]{SH16} A. Sheykhi, H. R. Salahi, A. Montakhab,
JHEP \textbf{1604}, 058 (2016) [arXiv:1603.00075[gr-qc]].






\bibitem[SHsh(17)]{SHsh(17)} A. Sheykhi, F. Shaker, Int. J.
Mod. Phys. D \textbf{26}, 1750050 (2017) [arXiv:1606.04364[gr-qc]].



\bibitem[SHSH(16)]{SHSH(16)} A. Sheykhi, F. Shaker, Can. J. of Phys.
\textbf{94}, 1372 (2016) [arXiv:1601.05817[hep-th]].



\bibitem{shSh(16)} A. Sheykhi, F. Shaker, Phys Lett. B \textbf{754}, 281
(2016) [arXiv:1601.04035[hep-th]].



\bibitem{n5} S. I. Kruglov [arXiv:1801.06905[hep-th]].



\bibitem{mahya} M. Mohammadi, A. Sheykhi and M. Kord Zangeneh, Eur. Phys. J.
C \textbf{78}, 654 (2018) [arXiv:1805.07377[hep-th]].
\bibitem[BCS(1957)]{BCS57} J. Bardeen, L. N. Cooper, J. R. Schrieer, Theory of Superconductivity, Phys.
Rev. \textbf{108}, 1175 (1957). 



\bibitem{superp} A.P. Mackenzie, Y. Maeno, Physica B \textbf{280}, 148 (2000).









\bibitem{Caip} R.-G. Cai, S. He, L. Li, and L.F. Li. JHEP,
{\bf1312}, 036 (2013).



\bibitem{cai13p} R.G.Cai, L. Li, L.F. Li, JHEP, \textbf{1401}, 032 (2014)
[arXiv:1309.4877[hep-th]]. 






\bibitem{Donos} A. Donos and J.P. Gauntlett, JHEP {\bf12}, 091 (2011).
\bibitem{Gubser} S. S. Gubser and S. S. Pufu, JHEP, {\bf0811}, 033 (2008).
\bibitem{chaturverdip15} P. Chaturvedi, G. Sengupta, JHEP, \textbf{1504},
001 (2015) [arXiv:1501.06998 [hep-th]].



\bibitem{Roberts8} M. M. Roberts and S. A. Hartnoll, JHEP \textbf{0808}, 035
(2008) [arXiv:0805.3898[hep-th]].



\bibitem{zeng11} H. B. Zeng, W. M. Sun and H. S. Zong, Phys. Rev. D \textbf{%
83}, 046010 (2011) [arXiv:1010.5039 [hep-th]].



\bibitem{cai11p} R. G. Cai, Z. Y. Nie and H. Q. Zhang, Phys. Rev. D \textbf{%
83}, 066013 (2011) [arXiv:1012.5559[hep-th]].



\bibitem{pando12} L. A. Pando Zayas and D. Reichmann, Phys. Rev. D \textbf{85%
}, 106012 (2012) [arXiv:1108.4022[hep-th]].



\bibitem{momeni12p} D. Momeni, N. Majd and R. Myrzakulov, Europhys. Lett.
\textbf{97}, 61001 (2012) [arXiv:1204.1246[hep-th]].




\bibitem{mahyap} M. Mohammadi, A. Sheykhi and M. Kord Zangeneh, Eur. Phys. J.
C \textbf{78}, 984 (2018) [arXiv:1901.10540[hep-th]].



\bibitem{francessco1} F. Benini, C. P. Herzog, R. Rahman and A. Yarom, JHEP \textbf{1011}, 137 (2010) [arXiv:1007.1981[hep-th]].
\bibitem{francessco2} F. Benini, C. P. Herzog and A. Yarom,
Phys. Lett. B \textbf{701}, 626 (2011) [arXiv:1006.0731[hep-th]].
\bibitem{mahyalif} M. Mohammadi, A. Sheykhi, Eur. Phys. J. C \textbf{80}, 928 (2020), [arXiv:2010.07105 [hep-th]].


\bibitem{4gb1} D. Glavan and C. Lin, Phys. Rev. Lett. \textbf{124}, 081301(2020) [arXiv:1905.03601[gr-qc]].


 \bibitem{4gb2} A. Kumar and R. Kumar [arXiv:2003.13104[gr-qc]].
\bibitem{4gb3}S. U. Islam, R. Kumar and S. G. Ghosh, JCAP \textbf{09}, 030 (2020) [arXiv:2004.01038[gr-qc]].
\bibitem{4gb4} X.-H. Jin, Y.-X. Gao and D.-J. Liu, Int. J. Mod. Phys. D \textbf{29}, 2050065 (2020) [arXiv:2004.02261[gr-qc]].
\bibitem{4gb5}S.-J. Yang, J.-J. Wan, J. Chen, J. Yang and Y.-Q. Wang, Eur. Phys. J. C \textbf{80}, 937 (2020) [arXiv:2004.07934[gr-qc]].
\bibitem{4gb6}S. Ying, Chin. Phys. C \textbf{44}, 125101 (2020) [arXiv:2004.09480 [gr-qc]].
\bibitem{4gb7}M. Guo and P.-C. Li, Eur. Phys. J C\textbf{80}, 588 (2020) [arXiv:2003.02523 [gr-qc]].
\bibitem{4gb8}S.-W. Wei and Y.-X. Liu, Eur. Phys. J. P. \textbf{136}, 436 (2021) [arXiv:2003.07769[gr-qc]].
\bibitem{4gb9}S. Devi, R. Roy and S. Chakrabarti, Eur. Phys. J. C \textbf{80}, 760 (2020)[arXiv:2004.14935 [gr-qc]].
\bibitem{4gb10}M.A. Cuyubamba, Phys. Dark Universe \textbf{31}, 100789 (2021) [arXiv:2004.09025[gr-qc]].
\bibitem{4gb11}P. Liu, C. Niu and C.-Y. Zhang, Chin. Phys. C \textbf{45}, 025104 (2021) [arXiv:2004.10620[gr-qc]].
\bibitem{4gb12}M. Gurses, T.c. Sisman and B. Tekin, Eur. Phys. J. C \textbf{80}, 647 (2020) [arXiv:2004.03390 [gr-qc]].
\bibitem{4gb13}M. Gurses, T.c. Sisman and B. Tekin, Phys. Rev. Lett. \textbf{125}, 149001 (2020) .[arXiv:2009.13508 [gr-qc]].
\bibitem{4gb14}K. Aoki, M.A. Gorji and S. Mukohyama, Phys. Lett. B \textbf{810}, 135843 (2020) [arXiv:2005.03859 [gr-qc]].
\bibitem{4gb15}K. Aoki, M.A. Gorji and S. Mukohyama, JCAP \textbf{09}, 014 (2020) [arXiv::2005.08428 [gr-qc]].

\bibitem{RBM1} R. A. Hennigar, D. Kubizňák, R. B. Mann and C. Pollack, JHEP \textbf{2020}, 27 (2020) [arXiv:2004.09472 [gr-qc]].

\bibitem{RBM2}R. A. Hennigar, D. Kubiznak, R. B. Mann, C. Pollack, Phys. Lett. B \textbf{808}, 135657 (2020) [arXiv:2004.12995 [gr-qc]].

\bibitem{hscgb4} X. Qiao, L. Ou Yang, D. Wang, Q. Pan, J. Jing, JHEP \textbf{12}, 192 (2020).
\bibitem{hscgb42}D. Ghoraia, S. Gangopadhyay, Phys. Lett. B \textbf{822}, 136699 (2021) [arXiv:2105.09423[hep-th]].
\bibitem{mahyagaussp} M. Mohammadi, A. Sheykhi, Eur. Phys. J. C \textbf{ 79}, 473 (2019), [arXiv:1908.07992[hep-th]].

\bibitem{caipp} R. G. Cai, Z. Y Nie and H. Q. Zhang, Phys. Rev. D \textbf{82}, 066007 (2010) [arXiv:1007.3321[hep-th]].
\bibitem{wen18} D. Wen, H. Yu, Q. Pan, K. Lin and W. L. Qian, Nucl. Phys. B
\textbf{930}, 255 (2018) [arXiv:1803.06942[hep-th]].
\bibitem{liu} Y. Liu, Q. Pan and B. Wang, Phys. Lett. B \textbf{702}, 94 (2011) [arXiv:1106.4353[hep-th]].
\bibitem{mahyap11} M. Mohammadi, A. Sheykhi, Phys. Rev. D \textbf{100}, 086012 (2019) [arXiv:1910.06082 [hep-th]].







\end{thebibliography}
\end{document}